\documentclass[10pt,journal]{IEEEtran}

\ifCLASSOPTIONcompsoc
   \usepackage[nocompress]{cite}
\else
\fi

\usepackage{url}
\usepackage{amsmath}
\usepackage{amsfonts}
\usepackage{graphicx}
\usepackage{algpseudocode}
\usepackage{algorithm}
\newtheorem{example}{Example}
\newtheorem{definition}{Definition}

\begin{document}

%\begin{frontmatter}

\title{Privacy Policy Negotiation in Social Media}


\author{Jose M. Such,
Michael Rovatsos% <-this % stops a space
\IEEEcompsocitemizethanks{\IEEEcompsocthanksitem Jose M. Such is with the School of Computing and Communications,
Lancaster University, Lancaster,  LA1 4WA, UK. \protect\\
e-mail: j.such@lancaster.ac.uk% <-this % stops a space
\IEEEcompsocthanksitem Michael Rovatsos is with School of Informatics, The University of Edinburgh,
Informatics Forum 2.12, 10 Crichton Street Edinburgh EH8 9AB, UK
e-mail: mrovatso@inf.ed.ac.uk.}% <-this % stops a space
\thanks{}}

\IEEEcompsoctitleabstractindextext{
\begin{abstract}
Social Media involve many shared items, such as photos, which may concern more than one user. The first challenge we address in this paper is to develop a way for users of such items to take a decision on to whom to share these items. This is not an easy problem, as users' privacy preferences for the same item may conflict, so an approach that just merges in some way the users' privacy preferences may provide unsatisfactory results. We propose a negotiation mechanism for users to agree on a compromise for the conflicts found. The second challenge we address in this paper relates to the exponential complexity of such a negotiation mechanism, which could make it too slow to be used in practice in a Social Media infrastructure. To address this, we propose heuristics that reduce the complexity of the negotiation mechanism and show how substantial benefits can be derived from the use of these heuristics through extensive experimental evaluation that compares the performance of the negotiation mechanism with and without these heuristics. Moreover, we show that one such heuristic makes the negotiation mechanism produce results fast enough to be used in actual Social Media infrastructures with near-optimal results.  
\end{abstract}

}

\maketitle

\IEEEdisplaynotcompsoctitleabstractindextext

\IEEEpeerreviewmaketitle

%\end{frontmatter}

\section{Introduction}

\IEEEPARstart{D}{espite} the unquestionable success of social media (Facebook recently achieved 1 billion users), privacy is still one of the major concerns with regards to these technologies \cite{gross2005information}. Moreover, this concern has even been increasing over the last few years because users are more aware of the privacy threats that social media entail \cite{stutzman2013silent}. Most social media users consistently criticise mainstream social media for providing very complex privacy controls. These are often too difficult to understand, require time-consuming manual configuration, and do not allow for appropriate privacy management. Users are required to set many privacy controls (Facebook has 61 privacy controls \cite{bonneau2010privacy}), they need to consider a huge space of possible accessors (the average Facebook user has more than 130 friends \cite{quercia2012personality}), and they may have to perform fine-grained modifications for many items (the average Facebook user uploads 22 photos\footnote{Facebook One Billion Fact Sheet \url{http://newsroom.fb.com/download-media/4227}}). %This actually turns out being even more complex, because users need to understand how their information will be shared in order to appropriately set privacy controls to comply with their privacy preferences. 
This makes most users unable to cope with the complexity of privacy management in social media, which has led to numerous incidents in which people have lost their jobs, have been cyberbullied, or have lost court cases due to the inappropriate communication of personal information through social media. Empirical evidence shows that this significantly discourages users to either join social media or to show high engagement when they join \cite{staddon2012privacy}, in terms of how much they participate in social media sites, e.g., the amount of photos they upload, the number of comments they post, etc. Indeed, the most common case is the latter: people usually join social media because they do not want to be left apart, but after they join they do not participate much because they are not able to manage their privacy in a satisfactory way.

To address this problem, new access control paradigms for social media have been recently proposed, such as relationship-based access control \cite{carminati2009enforcing,fong2011relationship,wang2012isac}. These new access control paradigms are aimed at better capturing the nature of information sharing in social media by considering users' relationships as a central concept. %In particular, access control models that consider users relationships are gaining momentum \cite{carminati2009enforcing,fong2011relationship,wang2012isac}. 
This is supported by many studies %developed in different research fields 
that provide evidence that user relationships are the main factor that drives human disclosure of personal information \cite{houghton2010privacy,wiese2011you,duck2007human,strahilevitz2005social}, and that they should play a crucial role when defining access control mechanisms for social media \cite{gates2007access}. In particular, privacy policies in relationship-based access control are neither defined based on individual persons nor on their roles, but on the relationships --- and specifically the strength of the relationships or intimacy --- that a user has to other users. 

%Recently, methods that help users to define privacy policies for SNSs have been proposed, such as relationship-based access control models for social networking services \cite{carminati2009enforcing, fong2011relationship}. 

The main limitation of state-of-the-art relationship-based access control models is that they only support single user decisions \cite{such13review}. That is, these proposals assume that only one user takes the decision of whether or not to grant access to an item. This user is usually the one who uploads the item or shares it in some other way. However, what should we do when the definition of a privacy policy involves more than one user? This is an issue that arises frequently, for example when photos depict different people so that the privacy preferences of all of them should be respected when deciding who should be able to view them. A solution to this problem is very challenging, mainly due to the fact that: (i) a simple solution would be to let users define their privacy preferences with respect to an item and then merge individual preferences into a joint disclosure policy, but there can be situations in which privacy preferences may conflict and it may not be obvious how to merge them; and (ii) detecting and solving conflicts manually can be very complex and time-consuming because of the number of possible shared items and the number of possible accessors to be considered by users.

In this paper, we propose the first automated method to detect privacy policy conflicts and resolve them using a negotiation mechanism. %In particular, we propose a negotiation mechanism to resolve the conflicts detected. 
The preferences that determine negotiation behaviour are based on the strength of the relationships among users. As proven by recent experiments \cite{wiese2011you}, and in line with state-of-the-art relationship-based access control mechanisms \cite{carminati2009enforcing,fong2011relationship,wang2012isac}, this is the most important factor that users consider when deciding what information to disclose. Moreover, our mechanism uses the well-studied \emph{one-step negotiation protocol} \cite{rosenschein94} and  strategies which are known to be complete, efficient and stable. 

The second main challenge we address in this paper relates to the inherent complexity of considering the space of all possible deals users may achieve to solve the conflicts so that the results are optimal. In particular, we show in this paper that this space grows exponentially in the number of conflicts that need to be negotiated for. This is very important because this could prevent the mechanism from being used in actual Social Media infrastructures, as it would be too slow for users to be able to run the mechanism in real-time when they are posting items in the particular Social Media infrastructure. 

We overcome the complexity problem by developing a number of suitable heuristics. The aim is to reduce the space of all possible deals users may achieve to only those \emph{most promising}, so that the complexity is reduced while the outputs of the negotiation mechanism remain near-optimal. Through an extensive experimental comparison of the performance of the negotiation mechanism with and without heuristics, we show that they provide a significant search space reduction while remaining near-optimal.
One particular heuristic is able to produce results very close to the optimal fast enough to be used in real-world social media.

%In this mechanism, the preferences that determine negotiation behaviour are based on the strength of the relationships among users. This is the most important factor that users consider when deciding the information to disclose to other users, as proven by recent experiments \cite{wiese2011you}, and in line with the new relationship-based access control mechanisms \cite{carminati2009enforcing,fong2011relationship}. Once conflicts are detected, our method uses the well-studied \emph{one-step negotiation protocol} \cite{rosenschein94} and  strategies which are known to be complete, efficient and stable. %To the best of our knowledge, relationship-based mechanisms to deal with privacy policy negotiation in SNSs have not been presented before. 
%Beyond the adaptation of a existing negotiation mechanism to this specific problem, 
%As the complexity of the problem is exponential (as we will show bellow), we also propose a modified version of the one-step protocol including a heuristic which helps reduce this complexity. % of the problem (which, as we will show below, is exponential). 
%Through an extensive experimental evaluation that compares the performance of the negotiation mechanism with and without this heuristic, we show that with this heuristic the negotiation mechanism is by far faster, while remaining almost complete and close to optimality, than without this heuristic. 

The remainder of the paper is structured as follows. Section \ref{sec:back} introduces the concept of intimacy and policies for relationship-based access control. Section 3 provides a brief overview of the mechanism. Section \ref{sec:detection} describes the method used to detect conflicts. Section \ref{sec:space} proposes the model for ranking possible negotiation outcomes based on existing empirical evidence. Section \ref{sec:protocol} describes the negotiation mechanism and its complexity. Section \ref{sec:heuristic} introduces the heuristics we propose to reduce the complexity of the problem. Section \ref{sec:results} presents the experiments we conducted and discusses the results obtained. Section \ref{sec:relwork} reviews the related literature. Finally, Section \ref{sec:conclusions} presents some concluding remarks and describes possible avenues for future work.

%Section \ref{sec:intimacy} presents the concept of intimacy and its relation to privacy policies. Section \ref{sec:space} defines both the negotiation space and the zone of agreement in this kind of problems. 
%Section \ref{sec:protocol} describes the negotiation mechanism. Section \ref{sec:heuristic} introduces the heuristic we propose to reduce the complexity of the problem. Section \ref{sec:results} presents the experiments we conducted and discusses the results obtained. Section \ref{sec:relwork} discusses related literature. Finally, Section \ref{sec:conclusions} presents some concluding remarks and future work. %Finally, Section \ref{sec:future} proposes some future works. 

\section{Background}
\label{sec:back}
We consider a set of agents $Ag=N\cup T$, where a pair of \emph{negotiating} agents $N=\{a,b\}$ negotiate whether they should grant a set of \emph{target} agents $T=\{i_1,\mathellipsis,i_n\}$ access to a particular item $it$. For simplicity and without loss of generality, we will consider only a negotiation for one item throughout this paper -- for example, a photo that depicts the two users which agents $a$ and $b$ are representing -- and hence, we do not include any additional notation for the item in question. The problem we are considering is how $a$ and $b$ can detect whether their individual privacy preferences for the item are conflicting\footnote{Note that we focus on detecting conflicts once we know the parties that co-own an item and have their individual privacy preferences for the item. We are, however, not proposing a method to automatically detect which items are co-owned and by whom they are co-owned. This is a different problem that is out of the scope of this paper.}, and if they are conflicting, how $a$ and $b$ can achieve an agreement on which agents in $T$ should be granted access to this item.

Negotiating agents have the individual privacy preferences of their users about the item --- i.e., to whom of their online friends users would like to share the item if they were to decide it unilaterally. Users could define their individual privacy preferences about the item using any of the access control models already proposed for Social Media\footnote{Our approach does not even need users to specify their individual privacy preferences for each and every item, users could specify their preferences for groups or categories of items, e.g., users could specify the same preferences for all the photos in a photo album.}. In this paper, we use relationship-based access control \cite{carminati2009enforcing,fong2011relationship,wang2012isac} %to represent individual privacy preferences 
because this type of access control has emerged as an appropriate way to capture the nature of individual sharing preferences in Social Media, and because it makes the link between individual preferences and our proposed mechanism more intuitive. However, other approaches to access control in Social Media --- such as the group-based access control models of mainstream Social Media infrastructures (like Facebook lists or Google+ circles), or (semi-)automated approaches like \cite{fang2010privacy,squicciarini2011a3p,danezis2009inferring} --- can also be used in conjunction with our proposed mechanism, as we will be pointing out at different points in this paper. %Note that the results obtained would be the same, as the individual access control model used is only the way in which users' express their individual privacy preferences, which are then given to the mechanism we propose in this paper as an input.

\vspace{-6pt}

\subsection{Intimacy}
In modelling how agents' preferences about disclosure are formed, we base our analysis on  the concept of {\em relationship strength} (or {\em intimacy}\footnote{Over the course of this article, we shall use relationship strength, tie strength, and intimacy as equivalents.}) between two persons \cite{granovetter1973strength}. This is because in the domain of social networks, there is strong evidence that the best predictor for individuals' willingness to share a particular item with another individual is how close/intimate their relationship is \cite{green06,strahilevitz2005social,houghton2010privacy,wiese2011you}. It is important to note that intimacy does not equal social distance, which is usually measured as the number of hops (friends) between two users that are not necessarily directly connected with each other. Instead, intimacy is the measure of the relationship strength between two directly connected friends. Intimacy can also be transitive under certain conditions \cite{white2002navigability}, so some particular non-directly connected friends may have non-zero intimacy. The rest of non-directly connected friends who do not meet the conditions to have a transitive intimacy would have no intimacy at all --- e.g., they would have an intimacy value of 0. %Note, however, that this does not impose any limitation %, so that they could have either an intimacy value (obtained by transitivity) or an absence of intimacy (which would mean an intimacy value of 0). 

We assume that we have available the intimacies among users who use the particular social media service in which our mechanism is to be deployed. This is a realistic assumption, as 
intimacies can be accurately estimated from content users have previously published, using tools that obtain intimacies automatically for Facebook \cite{gilbert2009predicting,fogues2013bff}; Twitter \cite{gilbert2012predicting}; and the like%\footnote{Note, however, that tools like BFF \cite{fogues2013bff} do not provide any way or mechanism to manage privacy, they are only used to retrieve the intimacy that users have to each other based on the information they can extract from users' profiles in Social Media.}
. Even if these tools are not used, users can be asked to self-report their intimacies to their friends, but this would obviously mean more burden on the users. We formally represent the intimacy of two agents as follows:
\begin{definition}
Given two agents $a,b\in Ag$, and a maximum integer intimacy value $\mathcal{Y}$, the {\em intimacy} between $a$ and $b$ is given as $int(a,b)$, where $int: Ag \times Ag \rightarrow [0,\mathcal{Y}]$.
\end{definition}

%We assume that agents know (or are able to obtain) intimacy values for any other agent. This does not imply that users have to define their intimacy to other agents, as this can be obtained automatically as shown by many studies that have proposed specific methods and tools to reliably calculate the intimacy based on information gathered automatically from a SNS. This has been successfully done for Facebook \cite{gilbert2009predicting,xiang2010modeling}, Twitter \cite{gilbert2012predicting}, and even with methods based on mining e-mails exchanged among users \cite{mccallum2007topic}.  For the purposes of this paper, we assume a function that returns the intimacy of two agents as follows:
%\begin{definition}
%Given two agents $a,b\in Ag$, and a maximum integer intimacy value $\mathcal{Y}$, the {\em intimacy} between $a$ and $b$ is given as $int(a,b)$, where $int: Ag \times Ag \rightarrow [0,\mathcal{Y}]$.
%\end{definition}

The maximum possible intimacy $\mathcal{Y}$ depends on the scale used by the particular methods/tools used to obtain intimacy. For example, in Gilbert and Karahalios' approach \cite{gilbert2009predicting} $\mathcal{Y}=100$, and  %[\citeyear{gilbert2009predicting}] approach, for example,  $\mathcal{Y}=100$, and 
in Fogu\'{e}s et al. \cite{fogues2013bff} $\mathcal{Y}=5$. 

\subsection{Relationship-based Policies}
Privacy policies in relationship-based access control \cite{carminati2009enforcing,wang2012isac} consider the strength of the relationships (or intimacy) that a user has to other social media users. In particular, privacy policies define an intimacy threshold that users must have with the user that defines the privacy policy to access the specific item. For instance, if a user defines a privacy policy with a high intimacy threshold, it means that only users who are very close to this user can access the item. Privacy policies in this approach also consider relationship types $R=\{r_1,\mathellipsis,r_n\}$ --- e.g., family, friends, colleagues --- so that a different intimacy threshold is defined for each relationship type. %Different relationship types $R=\{r_1,\mathellipsis,r_n\}$ are also considered, so that a different intimacy threshold is defined for each relationship type, e.g., family, friends, colleagues, etc. 
Moreover, a mapping $r: Ag\times Ag \rightarrow R$ is defined so that $r(a,b)$ is the relationship type between agents $a,b\in Ag$.

\begin{definition}
A privacy policy is a tuple $P=\langle \theta_1,\mathellipsis,\theta_{\mid R\mid},E\rangle$, where $\theta_j\in [0,\mathcal{Y}]$ is the intimacy threshold for the relationship type $r_j\in R$, and $E\subseteq T$ is the set of exceptions to the policy.
\end{definition}

%We assume that each agent $a\in Ag$ has a most preferred privacy policy $P_a$ for the item under negotiation, which describes the minimum intimacy that target agents in $T$ must have with $a$ to access the item, as well as if there is any particular exception. We also assume that each agent $a$ has two least preferred policies $L_a$ and $R_a$ for the item under negotiation, so that $L_a.thr\le P_a.thr\le R_a.thr$. This is because users may want not only avoid that some of their friends have access to an item but also ensure that some other \emph{closer} friends do have access to the item \cite{palen2003unpacking}. The least preferred policies therefore would act in a similar way to \emph{reservation prices} in electronic auctions. It is important to note at this point, that the particular method/tool used to elicit/define the privacy policies that match users' preferences is out of the scope of this paper\footnote{Privacy policies for social networking services can be obtained either in a (partially) automated way requiring little intervention from the user (e.g. by using active learning techniques \cite{fang2010privacy}, data mining and image classification techniques \cite{squicciarini2011a3p}, or by means of recommendations \cite{jakob2011content}), or in a way that requires complete intervention from the human user (such as in \cite{wishart2010collaborative}).}. Our proposal is agnostic regarding this, we only consider that privacy policies have been already defined as a previous step to detect conflicts and resolve them using the mechanisms proposed in this paper.

We denote $P_a$ as the preferred privacy policy of the user which agent $a\in Ag$ is representing. % in the negotiation process. 

\section{Mechanism Overview}
The mechanism proposed proposed in this paper takes as inputs the individual privacy policy of each negotiating agent --- $P_a$ for all $a\in N$ --- and the intimacies among agents --- $int(a,b)$ for all $a,b\in Ag$ ---, elicited as described in the previous section. The mechanism has two stages, as shown in Figure 1:
%\vspace{-0.1cm}
\begin{enumerate}
\itemsep0cm
\item The individual privacy policies of the negotiating agents are inspected to identify any conflict, as described in Section \ref{sec:detection}.
\item If conflicts are found, then agents run the negotiation mechanism described in sections \ref{sec:space} and \ref{sec:protocol} to resolve every conflict found. 
\end{enumerate}

\begin{figure}[!htb]
\centering
%\hspace{-10pt}
\vspace{-16pt}

\includegraphics[scale=0.229]{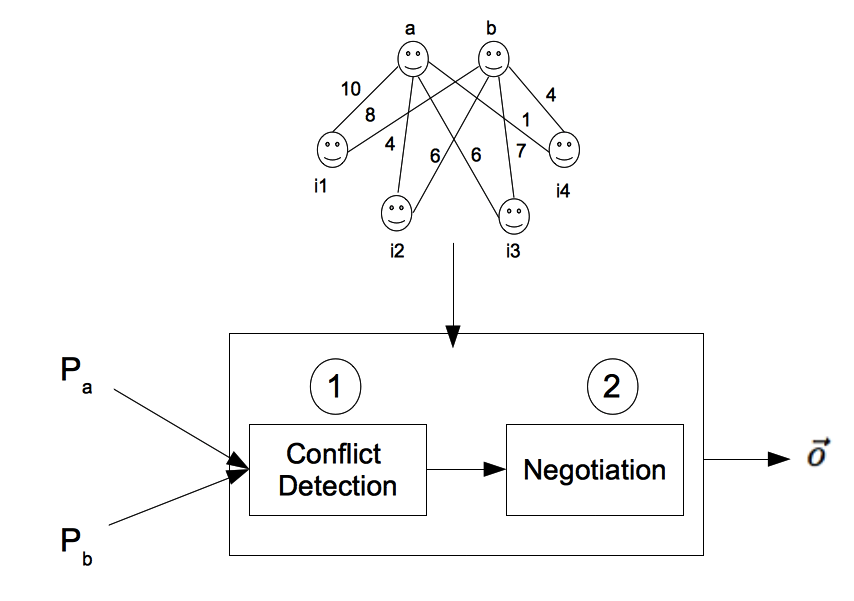}
\label{fig-mo}
\vspace{-0.5cm}
\caption{Mechanism Overview.}
\end{figure}

\vspace{-0.5cm}

\section{Privacy policy conflict detection}
\label{sec:detection}
In this section, we describe how conflicts between the preferred privacy policies defined by agents $a$ and $b$ on the item under consideration can be detected. Each agent usually has different intimacies to other agents, so that two privacy policies from two different agents can only be compared in terms of their effects. For instance, suppose that agents $a$ and $b$ have the same preferred privacy policies for a particular item (i.e., these policies have the same intimacy thresholds for the same relationship types). Suppose also that agents $a$ and $b$ have different intimacies to agent $i_1$ in $T$. If agent $a$ has an intimacy with $i_1$ below the corresponding threshold but agent $b$ has an intimacy with $i_1$ above the threshold, their individual decisions on whether to grant access to $i_1$ would be different. A similar example can be constructed for the case where the two policies are different but, because of the individual intimacies of agents $a$ and $b$ to other agents in $T$, they suggest the same decision. Thus, we need to consider the effects that each particular policy has on the set of target agents $T$ to determine whether or not the policies of two negotiating agents are in conflict. That is, we need to know which agents are granted/denied access by a given policy. 

Privacy policies dictate a particular action to be performed when an agent in $T$ tries to access the item. The available actions are usually either $0$ (denying access) or $1$ (granting access). The action to perform according to a given privacy policy is determined as follows:

%In order to determine whether or not the privacy policies of two negotiating agents are in conflict, we need to consider the effects that each particular privacy policy has on the set of target agents $T$. That is, we need to know which agents are granted/denied access by a privacy policy. Note that each user will have different intimacies to other users, so that two privacy policies from two different agents can only be compared in terms of their effects. For this, we consider that privacy policies dictate a particular action to be performed when an agent in $T$ will try to access the item. We assume that the available actions are either $0$ (denying access) or $1$ (granting access). The action to perform according to a given privacy policy is determined as follows: 

\begin{definition}
Given an agent $a\in Ag$, its privacy policy $P_a=\langle \theta_1,\mathellipsis,\theta_{\mid R\mid},E\rangle$, and an agent $i\in T$, we define the function\footnote{This function will be different depending on the access control model used. For instance, in group-based access control (like Facebook lists) this function could be specified as:
\begin{displaymath}
act(P_a,i) =  
\begin{cases}
%1 & \text{iff } int(a,i)\ge thr \wedge i\notin E \wedge P_a=\langle thr, E\rangle\\
1 & \text{iff $i$ is in a group that is granted access and}\\
&  \text{$i$ is not blocked individually} \\
1 & \text{iff $i$ is granted access individually} \\
%Yes & \text{if } Y_a(P_a) \leq y(a,c) \leq Y_a(L_a)\\
0 & \text{otherwise}
\end{cases}
\end{displaymath}%will return 1 if $i$ is in a group that is granted access to the item or if $i$ is granted access individually to the item. Otherwise, the action for $i$ would be 0.
}:

\vspace{-20pt}

\begin{displaymath}
act(P_a,i) =  
\begin{cases}
%1 & \text{iff } int(a,i)\ge thr \wedge i\notin E \wedge P_a=\langle thr, E\rangle\\
1 & \text{iff } int(a,i)\ge \theta_{r(a,i)} \wedge i\notin E\\
%Yes & \text{if } Y_a(P_a) \leq y(a,c) \leq Y_a(L_a)\\
0 & \text{otherwise}
\end{cases}
\end{displaymath}
\end{definition}
%Where $action(P_a,c)=1$ means that $P_a$ grants $c$ access to the item under negotiation, and $action(P_a,c)=0$ means that $P_a$ denies $c$ access to the item under negotiation.

We also consider so-called \emph{action vectors} $\vec{v}\in\{0,1\}^n$, i.e.\ complete assignments of actions to all agents in $T$, so that $v_i$ denotes the action for $i\in T$. When a privacy policy is applied to $T$, it produces such an action vector: 

%as a complete assignment of actions to all agents in $T$, so that we will denote $v_i$ as the component of the vector containing the action for agent $i\in T$. When a privacy policy is applied to the set of agents $T$, it produces such an action vector: %a privacy policy we extend the definition of the function $act$ to deal with action vectors as 

%\ct{, i.e.\ complete assignments} of actions to all agents in $T$, \ct{such that} $v_i$ \ct{denotes} the action for agent $i\in T$. When a privacy policy is applied to the set of agents $T$, it produces such an action vector:

%follows:

%We also define the concept of action vector, which is the vector that specifies for each agent the action that a privacy policy produces. Formally:

%From this, we also define the set of agents to which a privacy policy grants access as:

%\begin{definition}
%Given a privacy policy $P\in [0,\mathcal{Y}]$, the set of agents to which this policy grants access is $G^P=\{c\mid c\in Ag \wedge action(P,c)=grant\}$.  
%\end{definition}
\begin{definition}
The {\em action vector} induced by privacy policy $P$ in $T$ is $\vec{v}=(v_1,\mathellipsis,v_n)$, where $n=\mid T\mid$ and $v_i=act(P,i)$.  
\end{definition}
%\section{Problem Statement}
%\label{sec:statement}
%We assume a set of agents $Ag=\{a_1,\mathellipsis,a_N\}$ and a set of items $I=\{it_1,\mathellipsis,it_n\}$. Moreover, we define a privacy policy $P_{it}\subseteq Ag$ as a set of agents that are granted access to an item $it$. We assume that an agent $a\in Ag$ that is not in $P_{it}$ is not granted access to item $it$. For simplicity and without loss of generality, we will refer to the very same item over the course of this article. Thus, we do not include the subindex specifying the item to which the privacy policies are referring. 
%\section{Problem Statement}
%\label{sec:problem}

We now consider the following problem: agents $a$ and $b$ have their own privacy policies for the same item, and the effect of these policies leads to different action vectors, i.e., there are target agents who are granted access according to one policy but denied access according to the other. In this case, we say that these two privacy policies are {\em in conflict}:
\begin{definition}
Given agents $a$ and $b$, their preferred privacy policies for the item under negotiation $P_a$ and $P_b$, and the action vectors induced by $P_a$ and $P_b$, which are $\vec{v}$ and $\vec{w}$ respectively, we say that $P_a$ and $P_b$ {\em are in conflict} with respect to the item under negotiation iff $\vec{v}\neq \vec{w}$.
\end{definition}  %A BIT OF A TRIVIAL DEFINITION, ALREADY SAID IN WORDS.

Further, we say that the {\em agents in conflict} is the set $\mathcal{C}=\{i\in T \mid v_i\neq w_i\}$. The complexity of the conflict detection mechanism is $\mathcal{O}(m\cdot n)$, where $m=\mid T \mid$ is the number of target agents, and $n=\mid N \mid$ is the number of negotiating agents. 

 % --- i.e.~the set of agents to which the two privacy policies assign a different action. 
% ALSO, IN THIS DEFINITION WE SEE FOR THE FIRST TIME THE HORRORS OF USING TOO MANY SUPERSCRIPTS AND SUBSCRIPTS - TRY TO SIMPLIFY THE NOTATION IN THE WHOLE PAPER (E.G. CALL TWO POLICIES P AND Q, TWO VECTORS V AND V' ETC)
%\vspace{-5pt}
\begin{table}[h]
%\begin{footnotesize}
\begin{center}
\begin{tabular}{|c|c|c|c|c|} \hline
&$i_1$ & $i_2$ & $i_3$ &$i_4$ \\ \hline
a  &10 & 6 & 4 & 1 \\ \hline
b  & 8 & 6 & 7 & 4 \\ \hline
\end{tabular}
\caption{Intimacies for Example 1.}
\label{tab:int}
\end{center}
%\end{footnotesize}
\end{table}

%\vspace{-20pt}

\begin{example}
Suppose a set of agents $Ag=\{a,b,i_1,i_2,$ $i_3,i_4\}$ and a unique relationship type among them $R=\{r_1\}$. Agents $a$ and $b$ are to decide which agents to grant access to a photo in which both of them are depicted, and the intimacy values of agents $a$ and $b$ toward others are as shown in Table~\ref{tab:int}, with $\mathcal{Y}=10$. 
Suppose that agent $a$ would prefer the policy $P_a=\langle 5,\emptyset\rangle$, so that $\vec{v}=(1,1,0,0)$ --- i.e., agent $a$ wants to grant access  to agents $i_1$ and $i_2$, toward whom she has an intimacy greater or equal to 5, but not to agents $i_3$ and $i_4$ who are less intimate to her. However, agent $b$ would prefer the policy $P_b=\langle 4,\emptyset\rangle$, so that $\vec{w}=(1,1,1,1)$ --- i.e., agent $b$ wants to grant access to agents $i_1$, $i_2$, $i_3$, and $i_4$. As $\vec{v}\neq \vec{w}$, $P_a$ and $P_b$ are in conflict and the set of agents in conflict is $\mathcal{C}=\{i_3,i_4\}$.
\end{example}

%\vspace{-10pt}

%\vspace{-20pt}

%\vspace{-15pt}

%\subsection{Acceptable Outcomes}
%\label{sec:space}
\section{Deals and their Utility}
\label{sec:space}
When agents run into a conflict, they can still negotiate a common action vector for the item in question to achieve a compromise, even if this will not result in an optimal policy for either of them. Such an outcome (or deal) is simply an action vector $\vec{o} \in \{0,1\}^n$ such that $n=\mid T\mid$, and the negotiation space is the space of all such vectors, which agents can rank according to utility functions that compactly reflect agents' preferences as will be defined in this section. Based on these utility functions, agents will agree on a particular action vector following the negotiation mechanism that we present in Section~\ref{sec:protocol}. 

\vspace{-10pt}

\subsection{From Deals to Local Policies}
After agents agree on a particular action vector, they must represent it in the form of a local privacy policy, so that users could consult it without having to check who is granted access or not individually. %That is, the negotiating agents firstly negotiate a common action vector, and when they agree on this action vector, it is translated to an appropriate privacy policy for each of them. 
A particular action vector is likely to be represented with different privacy policies for each agent because each of them has different intimacies toward agents in $T$. Also, it is crucial that the resulting privacy policies are as simple (in terms of the number of exceptions they include) as possible to ensure that users can understand them, which is crucial for an appropriate privacy management \cite{cranor2005security}. For instance, a privacy policy that includes 100 exceptions will be far more difficult to read and understand by the user than a privacy policy that only includes one exception. Thus, we are interested in a privacy policy that minimises the number of exceptions among all the privacy policies that can induce the same action vector. %represent the same policy. - esta frase esta mal

\begin{definition}
Given an action vector $\vec{o}$, the privacy policy that induces $\vec{o}$ in $T$ and minimises the number of exceptions is defined as:
\vspace{-7pt}
\begin{displaymath}
P^{\vec{o}} =  \arg\min_{P=\langle \theta_1,\mathellipsis,\theta_{\mid R\mid},E\rangle, \ \vec{v}=\vec{o}} \mid E \mid
\end{displaymath}
\end{definition}

\vspace{-20pt}

\subsection{Utility Function}
%We define the utility of a particular outcome (action vector) $\vec{o}$  as the distance (in terms of intimacy) from it to the action vector induced by the most preferred privacy policy of an agent. 
Intuitively, the rationale of the utility function agents will use to rank possible negotiation outcomes is: %(i) an agent will only consider outcomes that would be acceptable for her/him, any other outcomes will be assigned zero utility; (ii)  acceptable 
\begin{enumerate}
\item An outcome will be ranked based on the distance (in terms of intimacy) between the agent's preferred privacy policy and the privacy policy that induces the outcome, establishing an intimacy-based ordering of the outcomes as suggested by \cite{wiese2011you}. That is, the farther the privacy policy that induces the outcome is from the agent's preferred privacy policy, the less valued the outcome will be. 
\item An outcome will be ranked according to the number of exceptions of the privacy policy that induces the outcome, so that privacy policies should include as few exceptions as possible to ensure readability and understandability. That is, the more exceptions the privacy policy that induces the outcome entails, the less valued the outcome will be. 
\end{enumerate}
%We define the utility of a particular outcome (action vector) $\vec{o}$ as its distance (in terms of intimacy) from the action vector induced by the agent's preferred privacy policy. 
%Intuitively, the rationale of the utility function presented below is based on %(i) an agent will only consider outcomes that would be acceptable for her/him, any other outcomes will be assigned zero utility; (ii)  acceptable 
%establishing an intimacy-based ordering as suggested by \cite{wiese2011you} while minimising the number of exceptions as suggested by \cite{cranor2005security}. 

We start defining the intimacy distance between two policies. Privacy policies may have different intimacy dimensions (i.e., one per each relationship type considered), so that a metric in the $\mathbb{R}^{\mid R \mid}$ space is needed to compare them. We use the Euclidean distance to measure the distance between two policies as follows:
\begin{definition}
Given two policies $P$ and $Q$ the distance between them is\footnote{An example of this function for group-based access control models could be the euclidean distance but considering, for each possible group (instead of for each possible relationship type), the distance between the minimum intimacy of the users in this group that are granted access in the two policies compared (instead of the difference between intimacy thresholds).}:
\vspace{-10pt}
\begin{displaymath}
d(P,Q) =  
%\frac{1}{\mid R\mid}\cdot\sum_{r\in R} \mid P.\theta_r - Q.\theta_r \mid
%\sqrt{\sum_{r\in R} ( P.\theta_r - Q.\theta_r )2}
\sqrt{\sum_{r\in R} (P.\theta_r - Q.\theta_r)^2}
\end{displaymath}
\end{definition}

The advantage of using Euclidean distance is that it is sensitive to large variations in one dimension (relationship type). For instance, for two policies with a large difference in the ``friends'' relationship type (e.g., one of them grants any acquaintances access and the other only grants access to close friends) the distance would be large as well. This aligns with empirical evidence that suggests that intimacy distance plays a significant role on deciding to disclose or not \cite{green06,strahilevitz2005social,houghton2010privacy,wiese2011you}, so the higher the intimacy distance between what the user would like for a particular relationship type and the intimacy between that user and a target user, the less the user would like to share with the target use. 

Other metrics that can be used to compare vectors in the same space, such as the average of differences for each relationship type, the Chebyshev distance or the Manhattan distance do not always align with that. A simple average of the differences of intimacy in each relationship type, would lead to the final distance value being significantly attenuated if the differences in other relationship types (e.g., ``family'',``colleagues'', etc) are low. The Chebyshev distance is the maximum of all distances in each dimension. This would work well for the case in which there is a large variation on only one dimension (which would be the maximum), but if there are large variations in more than one dimension, this would not be accounted as only the maximum would be considered. In contrast, the Euclidean distance would clearly pick this difference, so that the final distance would be higher for the case in which there are large variations in more than one dimension. Finally, the Manhattan distance is the sum of the differences for all dimensions. Clearly, we could have small to medium variations in some dimensions that would add up to the same value as if only one dimension had a large variation. For instance, small variations in relationship types friends, family, colleagues, etc, could give the same result that a large variation in only one dimension (grant access to all your work colleagues instead of only to those you are closer to you), which would be clearly worse in terms of privacy implications. Therefore, the intimacy distance would be unable to catch these nuances, which might impact on how the utility function is able to model user preferences in this domain.

%d(P_a, P^{\vec{o}}) is the euclidean distance between the preferred privacy policy of agent $a$ and the privacy policy that induces $\vec{o}$ in $T$

%That is, we compare two policies based on the euclidean distance to account for the difference between intimacy thresholds for each relationship type\footenote{Note that by using the euclidean distance we avoid that high variations in one dimension could be hided }. %Note that we assume that relationship types are equally important. 
We now define the utility function based on the privacy policy that minimises the number of exceptions among all the privacy policies that induce the same action vector (Def. 6) and the intimacy distance (Def. 7).

\begin{definition}
Given agent $a$ and its preferred privacy policy $P_a$, the utility of an action vector $\vec{o}$ for agent $a$ is:
\vspace{-5pt}

\begin{displaymath}
u_a(\vec{o}) =  
%\begin{cases}
\lambda_a(P^{\vec{o}}) \cdot (\mathcal{D} - d(P_a, P^{\vec{o}}) )    %& \text{iff } acc(a,\vec{o})\\
%0 & \text{otherwise}
%\end{cases}
\end{displaymath}
\label{def:utility}
\end{definition}

\vspace{-5pt}

In this equation, $\mathcal{D}$ accounts for the maximum possible distance between two privacy policies, which would be obtained if the difference wss $\mathcal{Y}$ for all the relationship types, and

\vspace{-5pt}

\begin{displaymath}\lambda_a(P^{\vec{o}})=1-\frac{\mid P^{\vec{o}}.E\mid}{\mid T \mid}\end{displaymath} 

\noindent accounts for the number of exceptions that $P^{\vec{o}}$ --- the privacy policy that induces $\vec{o}$ in $T$ with the minimum possible exceptions, as defined above %in section \ref{sec:space} 
--- would entail (denoted as $\mid P^{\vec{o}}.E\mid$) with respect to the maximum number of exceptions possible, i.e., the number of target agents $\mid T \mid$. %\mathbb{E}_a$ -- the maximum number of exceptions entailed by any of the acceptable outcomes for agent $a$,  defined as \begin{displaymath}\mathbb{E}_a=\max_{\vec{x}%\mid \text{acc}(a,\vec{x})
%} \mid P^{\vec{x}}.E \mid\quad . \end{displaymath} 

Thus, although the utility of an outcome $\vec{o}$ is mainly based on intimacy distance between policies $P_a$ and $P^{\vec{o}}$, it is also weighted by a factor depending on the number of exceptions that the resulting privacy policy $P^{\vec{o}}$ would entail.

\section{Negotiation Protocol}
\label{sec:protocol}
Next, we consider how a mutually acceptable action vector $\vec{o}$ can be agreed upon by two negotiating agents $a,b\in N$ for the particular item under negotiation. In order for $a$ and $b$ to be able to negotiate a common action vector for a given item, we need to define a negotiation mechanism. A negotiation mechanism is composed of: (i) a negotiation protocol, which is a means of standardising the communication between participants in the negotiation process by defining how the actors can interact with each other; and (ii) strategies that agents can play over the course of a negotiation protocol \cite{rosenschein94}. %--- We assume that a deal $P\subseteq Ag$ is a privacy policy that includes all the agents $c\in Ag$ to be granted access.
Negotiation protocols determine the strategies agents can play during the execution of the negotiation protocol. Although there are some negotiation protocols proposed in the related literature \cite{lopes2008negotiation}, not all of them comply with the requirements for the domain we are tackling in terms of the strategies they permit. In particular, the requirements are: 
\begin{itemize}
\item The protocol must permit negotiation strategies that are stable. A stable strategy is one whereby if one agent is playing it the others' best strategy is to also play the same strategy. This is very important, as it is expected that each agent usually cares only about his user's own utility and will always try to play the strategy that can get the highest utility, even if it means everyone else is much worse \cite{vidal2010}. This could clearly lead to outcomes in which one agent gets an extremely high utility and everyone else gets almost nothing, which would be unfair. Therefore, stability is a very desirable property for our domain, as the negotiation mechanism should be fair, so that one user cannot just impose her preferences on the others.

\item The protocol must permit negotiation strategies that converge to the optimal solution (what is know in the negotiation literature as \emph{efficiency}). This is of crucial importance because of two main reasons: (i) if the negotiation protocol does not allow negotiation strategies that make the negotiation converge, agents could keep negotiating forever; (ii) the preferences of all the negotiating agents' users should be respected as much as possible. 
\end{itemize}

The simplest negotiation protocol that has these properties is the \emph{one-step protocol} \cite{rosenschein94}. This protocol has only one round, where the two agents propose a deal, and they must accept the deal that maximises the product of both agents' utilities. In case deals have the same product, one of them is chosen randomly (e.g.~by flipping a coin). In our case, a deal is an action vector. Thus, each agent will propose an action vector and they will accept the one that maximises the product of their utilities. To calculate individual utilities, agents  use the utility function presented in Section 4, which, in turn, uses the agents' preferred privacy policies elicited to detect conflicts (as explained in Section 3), and agents' intimacies, which can be accurately estimated from content already published (as explained in Section 2).

It was formally proven that the best strategy that agents can follow in this protocol is to propose the deal (action vector) that is best for themselves amongst those with maximal product of utilities. This strategy is both stable and efficient \cite{rosenschein94}. Its stability derives from being a Nash equilibrium (if one of the two agents follows this strategy the other agent's best strategy is to also follow this strategy), no agent has anything to gain by changing only his own strategy unilaterally. It is efficient in the sense that if there exists a solution % that is different from the conflict deal --- i.e. if there is an action vector that is acceptable for both agents ---, 
the agents will find it using this strategy. Other negotiation protocols such as the well-known monotonic concession protocol \cite{rosenschein94} or the alternating offers protocol \cite{osborne1990bargaining} do not allow strategies for this domain that would be both stable and efficient. The monotonic concession protocol is known to only allow strategies that can be either stable or efficient but not both \cite{endriss2006monotonic}. The alternating offers protocol has no convergence guarantees in its basic form, and for the time-dependent form utilities must be time-dependent \cite{vidal2010}, which is not the case in this domain. 

\begin{table}[h]
%\begin{footnotesize}
\begin{center}
\begin{tabular}{|c|c|c|c|} \hline
$\vec{o}$&$u_a(\vec{o})$ & $u_b(\vec{o})$ & $u_a(\vec{o})\times u_b(\vec{o})$ \\ \hline
(1,1,0,0)  &10 & 5 & 50 \\ \hline
(1,1,0,1)  &7.5 & 7.5 & 56.25 \\ \hline
(1,1,1,0)  & 9 & 7.5 & 67.5 \\ \hline
(1,1,1,1)  & 6 & 10 & 60 \\ \hline
\end{tabular}
\caption{Action vector utilities for Example 2.}
\label{tab:neg}
\end{center}
\vspace{-25pt}

%\end{footnotesize}
\end{table}
%\vspace{-15pt}

\begin{example}
Consider the application of our negotiation mechanism to resolve the conflicts detected in Example 1. As the set of agents in conflict was $\mathcal{C}=\{i_3,i_4\}$, the possible action vectors that could be selected as a compromise  are the ones in Table \ref{tab:neg} under the $\vec{o}$ column. Agents $a$ and $b$ will be able to rank each of the possible action vectors according to their preferences by measuring the utility of the action vectors, using the utility function from Definition \ref{def:utility}, which is calculated based on the preferred privacy policies of each agent and the intimacies between the negotiating agents and other target agents (shown in Table \ref{tab:int}). Table \ref{tab:neg} shows, for each action vector, the utilities for each individual agent as well as the product of both agents' utilities. For instance, the utility of $\vec{o}=(1,1,1,0)$ for $a$ is $u_a(\vec{o})=\lambda_a(\vec{o}) \cdot (\mathcal{D} - d(P_a, P^{\vec{o}}))$, where:
\begin{itemize}
\item $\mathcal{D}=10$ because we only have one relationship type in this example, so the maximum distance possible will be the maximum possible intimacy value $\mathcal{Y}=10$ in our example. 
\item $P^{\vec{o}}=\langle 4,\emptyset\rangle$ because this is the privacy policy that represents the action vector $\vec{o}=(1,1,1,0)$ for agent $a$ with the minimum number of exceptions. That is, all target agents with an intimacy higher or equal than 4 are granted access to the item under consideration.
\item $\lambda_a(\vec{o})=1$, because $P^{\vec{o}}$ does not entail any exception.
\item $d(P_a, P^{\vec{o}})=\sqrt{(P_a.\theta - P^{\vec{o}}.\theta)^2}=\sqrt{(5 - 4)^2}=1$.
\end{itemize} 
Thus, $u_a(\vec{o}=(1,1,1,0))=1 \times (10 - 1)=9$. Finally, agents will propose the action vector that is most favourable to them in terms of maximising the utility product. In our example, both will propose $\vec{o}=(1,1,1,0)$, which will be the outcome of the negotiation. 
\end{example}

\subsection{Complexity}
The number of possible deals in this setting is exponential in the number of agents in  $\mathcal{C}$, the set of agents in conflict. This is because the negotiation mechanism will have to consider $2^{\mid\mathcal{C}\mid}$ possible outcomes (action vectors) in order to find the one that maximises the product of utilities. However, the number of agents in $\mathcal{C}$ will change from negotiation to negotiation --- it completely depends on the preferred policies of the negotiating agents $a$ and $b$, so that we cannot predict the number of action vectors to be considered for each particular case \emph{a priori}, even though it seems clear that the more target agents in $T$ the more possibilities for the number of agents in $\mathcal{C}$ to be higher. In the worst case, all of the target agents $T$ will be in $\mathcal{C}$, i.e.\ the problem complexity grows exponentially in the number of target agents. Moreover, each possible outcome needs to be evaluated from the point of view of the other negotiating agents. Therefore, the complexity of the negotiation mechanism would be $\mathcal{O}(2^{l}\cdot n)$, where $l=\mid\mathcal{C}\mid$ is the number of conflicts, and $n=\mid N \mid$ is the number of negotiating agents. Finally, the worst case would be an upper bound for the complexity, so that in the worst case the complexity would be $\mathcal{O}(2^{m}\cdot n)$, where $m=\mid T \mid$ is the number of target agents, and $n=\mid N \mid$ is the number of negotiating agents.

\section{Heuristics}
\label{sec:heuristic}
To tackle the exponential blowup in the number of possible deals, we firstly considered \emph{complete} approaches that would decrease the complexity while always finding the \emph{optimal} solution. In particular, we considered approaches such as dynamic programming (and other divide and conquer algorithms) or branch and bound algorithms. However, none of these approaches was suitable for this domain. This is because the problem we are tackling does not exhibit the overlapping subproblems and optimal substructure properties required for a dynamic programming approach, and we were not able to develop a complete BnB algorithm because we were unable to find good-enough upper bounds in this domain for the utility of partial action vectors. For instance, using the utility function defined in Section 5 while ignoring target agents for whom no decision has been made (the one used for the Greedy heuristic) turned out to be too optimistic to prune enough nodes from the search space, so that the resulting BnB algorithm was not efficient enough to be used in practice. %Nonetheless, we followed some of the BnB principles to develop one of our heuristics presented below, but of course, as a heuristic this approach is incomplete --- there are no guarantees that the \emph{optimal} solution will always be found. 

As \emph{complete} approaches were not possible, we decided to develop heuristics --- i.e., \emph{incomplete} approaches --- that could reduce the number of action vectors considered when maximising the utility product to only those that appear most promising. This obviously involves a risk of losing optimality --- we empirically prove later on in Section \ref{sec:results} that these optimality losses remain within acceptable levels in practice.
%We propose to modify the protocol presented in the previous section by using heuristics to reduce the number of action vectors considered when maximising the utility product to only those that appear most promising. %This obviously involves a risk of losing optimality --- we empirically prove later on in Section \ref{sec:results} that the protocol with heuristics is by far faster and that the loses in optimality are acceptable.

\subsection{Distance-based Heuristic}
Our first heuristic fixes the action to be taken for some conflicting target agents without trying both possible actions (granting or denying access) when generating the possible action vectors. %Our heuristic is \dt{applied} \ct{local} in the sense that it \ct{considers only one target agent at a time while ignoring other parts of the action vector.} %Thus, it does not consider the utility of a whole action vector, which will imply considering the rest of target agentsi.e., it does not consider either the privacy policy distance or the number of exceptions induced when producing a policy from the resulting action vector. 

\begin{algorithm}[!h]
\begin{footnotesize}
\begin{algorithmic}[1]
\Require $T$,\ $\vec{v}$,\ $\vec{w}$,\ $P_a$,\ $P_b$,\ $\varphi$ \;
\Ensure $\vec{o}$
%\State
\State $maxval \gets 0$ 
\State $initialize\_to\_zeros(\vec{o})$
\State $initialize\_to\_zeros(\vec{t})$
%\State
\ForAll{$i\in T$} \Comment{Generating a partial action vector}
\If {$v[i] = w[i]$ }
    \State $t[i]\gets v[i]$
\Else 
        \Comment{Conflict - applying heuristic}
        			\State $d_a\gets\mid P_a.\theta_{r(a,i)}-int(a,i)\mid$
        			\State $d_b\gets \mid P_b.\theta_{r(b,i)}-int(b,i)\mid$
        			%\State
        			\If {$\mid d_a - d_b \mid \geq \varphi$}
        			%\State
        				\If {$d_a > d_b$}
	      					\State $t[i]\gets v[i]$
        				\Else 
        					\State $t[i]\gets w[i]$
        				\EndIf
        			\Else
        			\State $t[i]\gets *$  \Comment{Both will be considered}
        			\EndIf
    %    	\EndIf
   % \EndIf
\EndIf
\EndFor
%\State
\ForAll{$\vec{x}\in \mathcal{X}^{\vec{t}}$}   \Comment{Utility product maximisation}
\State $prod\gets u_a(\vec{x})\times u_b(\vec{x})$
\If {$prod > max$}
    \State $\vec{o}\gets \vec{x}$
    \State $max \gets prod$ 
    \State $maxUT \gets u_a(\vec{x})$
\EndIf
\If {$prod = max$ \textbf{and} $u_a(\vec{x}) > maxUT$} 
    \State $\vec{o}\gets \vec{x}$
    \State $maxUT \gets u_a(\vec{x})$
\EndIf
\EndFor
\end{algorithmic}
\end{footnotesize}
\caption{Distance-based Heuristic - Agent a}
\label{al:pg1}
\end{algorithm}

Informally speaking, the heuristic calculates how \emph{important} assigning a particular action to a particular target agent is for one of the negotiating agents. This importance is calculated by measuring the intimacy distance between a target agent and the threshold for the relationship type of this agent in the preferred privacy policies of both agents $a$ and $b$. If the difference between these two distances is higher than a threshold value $\varphi$ --- the so-called \emph{importance threshold} --- we consider that the action for that target agent is more important to the negotiating agent with higher distance than to the negotiating agent with the lower distance. Thus, we only consider action vectors in which the particular target agent is assigned the action suggested by the negotiating agent with the highest distance. % fix in the out the action suggested by the agent with the highest distance agent's action vector to the corresponding element in the action vector template.
Both negotiating agents are using the same $\varphi$ value.  
%\vspace{-5pt}

%\vspace{-20pt}
%\vspace{-5pt}

Algorithm \ref{al:pg1} illustrates the proposal generation process with the heuristic for agent $a$ (the pseudocode for agent $b$ would be the same except for lines 27-30 as explained below). In this algorithm, we use a so-called \emph{partial action vector} $\vec{t}$ with $t_i\in \{0,1,*\}$, where $0$/$1$ means that action vectors generated from that partial action vector will deny/grant target agent $i$ access and $*$ means that both actions will be considered when generating action vectors. For each target agent $i$, if both action vectors $\vec{v}$ and $\vec{w}$ assign the same action --- i.e. there is no conflict in the action to be taken for this particular agent, then this action is chosen for the partial action vector. If the action vectors do not assign the same action, we have a conflict. In case of conflict, we first measure the distances in intimacy of agent $i$ to both agents $a$ and $b$ (Lines 11 and 12). If the difference between these two distances is greater than or equal to the \emph{importance threshold} $\varphi$, we apply the action corresponding to the greater distance. If not, we assign a $*$ value to consider both alternatives. In section \ref{sec:results}, we discuss the effect of different \emph{importance thresholds} $\varphi$ on the performance of the negotiation mechanism. 

After the creation of the partial action vector, we only consider action vectors that comply with the partial action vector in the utility product maximisation step (from Line  21). This set is defined as $\mathcal{X}^{\vec{t}}=\{\vec{x}\mid \forall i\in T,\ x_i=t_i \vee t_i=* \}$. That is, we are constraining the set of action vectors considered when maximising the product of utilities for both agents. In Lines 22-26, if the product of utilities for the current action vector is higher than for the best action vector seen so far, agent $a$ updates the latter with the current one. In Lines 27-30, if the product of utilities for the current action vector is equal to the product of utilities of the best action vector seen so far and the individual utility for agent $a$ is higher for the current action vector than for the best action vector seen so far, agent $a$ updates the best action action vector with the current one. This is because, as explained in the previous section, the best strategy that agents can follow in this protocol is to propose the deal (action vector) that is best for themselves among those with a maximal product of utilities. Lines 26, 28, and 30 will change for agent $b$ to consider its individual utility instead of the individual utility of agent $a$.

\subsection{Greedy Heuristic}
Our second heuristic follows a greedy approach to propose an action (granting/denying access) for each conflict separately. The informal idea is to make the locally optimal choice at each stage with the hope of finding a global optimum. %The choice of which conflict to consider is based on the loss of total utility incurred by assigning an action to a conflict, as we will explain in more detail below. %We assume that both negotiating agents are using the heuristic to generate their proposals. 

Like the previous heuristic, this one also considers partial action vectors. Here, however, agents also calculate the utility of partial action vectors, because the greedy heuristic needs an estimation of how good the partial action vectors generated in each step of the heuristic are. To this aim, the utility function defined in Section \ref{sec:space} is used while ignoring target agents for whom no decision has been made so far, i.e.\ where the partial action vector contains $*$ at the entry corresponding to the respective target agent. This is done by considering that the action assigned to them is the most desired by each negotiating agent. In this way, the utility of a partial action vector acts as an upper bound of the utility that will be achieved when having a complete action vector. %\at{Isn't it a lower bound because the product is not incremented with a positive value at that position?}

%\vspace{-5pt}
\begin{algorithm}[!h]
\begin{footnotesize}
\begin{algorithmic}[1]
\Require $T$,\ $\vec{v}$,\ $\vec{w}$,\ $P_a$,\ $P_b$ \;
\Ensure $\vec{o}$
%\State
\ForAll{$i\in T$} \Comment{Detecting Conflicts}
\If {$v[i] = w[i]$ }
    \State $o[i]\gets v[i]$
\Else
    \State $o[i]\gets *$  
    \State $\mathcal{C}\gets \mathcal{C} \cup \{i\}$
\EndIf
 
\EndFor
\State %nConflicts $\gets \mid T\mid$
\While {$\mathcal{C}\neq \emptyset$} 
\State $maxVal \gets 0$ 
\State $\vec{x}\gets\vec{o}$ 

\ForAll{$i \in \mathcal{C}$}\Comment{For all the remaining conflicts}
%\State $prod\gets u_a(\vec{x})\times u_b(\vec{x})$
%\If {$x_i = *$}
	\For{$action \in \{0,1\}$}
	\State $x[i]\gets action$
    \State $prod=u_a(\vec{x})\times u_b(\vec{x})$ 
    \If {$prod > max$}  		
    		\State $max \gets prod$ 
    		\State $maxUT \gets u_a(\vec{x})$ 
    		\State $maxTarget\gets i$
    		\State $maxAction\gets x[i]$
	\EndIf
	\If {$prod = max$ \textbf{and} $u_a(\vec{x}) > maxUT$}  		
    		\State $maxUT \gets u_a(\vec{x})$ 
    		\State $maxTarget\gets i$
    		\State $maxAction\gets x[i]$
	\EndIf
	\EndFor
	\State $x[i]\gets *$
%\EndIf
\EndFor
\State $o[maxTarget]=maxAction$ %\Comment{The conflict that causes the least utility loss is solved}
\State $\mathcal{C}\gets\mathcal{C}\setminus \{maxTarget\}$
\EndWhile
\end{algorithmic}
\end{footnotesize}
\caption{Greedy Heuristic - Agent a}
\label{al:pg2}
\end{algorithm}

The heuristic starts by considering the action vector in which all agents not in conflict are assigned the corresponding (commonly agreeable) action, and all other agents are assigned a $*$ value. Thus, this first partial action vector always has the maximum product of utilities possible with respect to the other possible partial action vectors and complete action vectors. This is because if we ignore all conflicts, the actions assigned to the target agents will completely comply with the preferred privacy policies of the negotiating agents.

After this, the heuristic incrementally assigns an action to one conflict at a time. The choice of the conflict and the action taken to resolve it is made as follows: the heuristic compares all possible grant/deny configurations over all conflicts and greedily chooses the most promising option, i.e.\ the one that decreases the total product of utilities by the smallest amount. This process is repeated until all conflicts are resolved and a complete action vector has been produced.

Algorithm \ref{al:pg2} shows the pseudocode for the greedy heuristic for agent $a$ (the pseudocode for agent $b$ would be the same except lines 19, 23, and 24 would change for agent $b$ to consider its individual utility instead of the individual utility of agent $a$). We first detect conflicts and create a partial action vector by assigning either the corresponding action to the agents not in conflict or $*$ to the agents that are in conflict (lines 1-8). Then, while there are conflicts that have not been dealt with (line 10), we assign an action to a conflict, one conflict at a time. To achieve this, we try all possible partial action vectors by exploring 0 and 1 values for each of the conflicts (lines 13-24). Then, we select the conflict and action that maximise the product of utilities. Finally, we update the partial action vector with the selected action for the conflict in question and mark that conflict as resolved (lines 31-32). %in the pto resolve the conflict that will decrease the product of utilities to the least possible extent, and update the outcome template accordingly (Line 25). 

\subsection{GreedyBnB Heuristic}
Our third heuristic is GreedyBnB and it is loosely based on Branch and Bound (BnB) algorithms. These algorithms systematically discard large subsets of fruitless candidates from the search space by means of upper and lower estimated bounds of the quantity being optimised. Our GreedyBnB heuristic operates in a similar way to a BnB algorithm\footnote{As explained above, a complete BnB algorithm was not implemented because we were unable to find good-enough upper bounds in this domain for the utility of partial action vectors. Nonetheless, we followed some of the BnB principles to develop the GreedyBnB heuristic, but of course, as a heuristic this approach is incomplete --- there are no guarantees that the \emph{optimal} solution will always be found.}.
%For instance, using the utility function defined in Section \ref{sec:space} while ignoring target agents for whom no decision has been made (the one used for the Greedy heuristic) turned out to be too optimistic to prune enough nodes from the search space, so that the resulting BnB algorithm was not efficient enough to be used in practice.}. %, but it is not complete, i.e., it might not find the globally optimal solution. 
%However, as opposed to it is not complete (neither are the other aforementioned heuristics). In this case, we could not implement a complete BnB algorithm because we were unable to find good-enough upper bounds in this domain for the utility of partial action vectors. %, so a complete BnB algorithm was not efficient enough to be used in practice. 
%For instance, using the utility function defined in Section \ref{sec:space} while ignoring target agents for whom no decision has been made (the one used for the Greedy heuristic) turned out to be too optimistic to prune enough nodes from the search space, so that the resulting BnB algorithm was not efficient enough to be used in practice. %Therefore, we only consider lower bounds as the greedy solution obtained for partial action vectors. 
Informally speaking, the GreedyBnB heuristic explores other branches of the search space different from the one followed by the greedy heuristic. The aim is to explore branches of the search space that were discarded by the greedy heuristic at early stages and that could lead to better outcomes. The GreedyBnB heuristic uses the greedy heuristic as a selection algorithm to prioritise options. In particular, our GreedyBnB uses the Greedy heuristic to estimate a lower bound for the utility of partial action vectors. That is, given a partial action vector, we estimate its utility to be, at least, the utility of the solution obtained when the greedy heuristic is applied to that partial action vector. %Informally speaking, the GreedyBnB heuristic explores other branches of the search space different from the one originally chosen by the greedy heuristic. The aim is to explore branches of the search space that were discarded by the greedy heuristic at early stages and that could lead to better outcomes. 
 %, it follows the greedy heuristic to obtain a solution, so that the utility of this partial action vector is estimated to be at least the utility of the solution obtained. which of course might not be optimal, but that will always be one so that no other possible solution will be worse given the same partial action vector. Note, however, that given a partial action vector the greedy heuristic may not achieve the optimal solution. A known issue with greedy heuristics is that they take decisions 

% For example, all known greedy coloring algorithms for the graph coloring problem and all other NP-complete problems do not consistently find optimum solutions. Nevertheless, they are useful because they are quick to think up and often give good approximations to the optimum.

\begin{algorithm}[!h]
\begin{footnotesize}
\begin{algorithmic}[1]
\Require $T$,\ $\vec{v}$,\ $\vec{w}$,\ $P_a$,\ $P_b$ \;
\Ensure $\vec{o}$
%\State
\ForAll{$i\in T$} \Comment{Detecting Conflicts}
\If {$v[i] = w[i]$ }
    \State $t[i]\gets v[i]$
\Else
    \State $t[i]\gets *$  
%    \State $\mathcal{C}\gets \mathcal{C} \cup \{i\}$
\EndIf
 
\EndFor
\State %nConflicts $\gets \mid T\mid$

\State $\{\vec{o},maxVal\}=greedySolution(\vec{t}\ )$ \Comment{Obtain a solution with greedy heuristic}
\State $L.add(\{\vec{t},\vec{o},maxVal\})$
\State
\While {$L$ is not empty} \Comment{Explore other possible solutions}
\State $\{\vec{x},\vec{y},ut\} \gets L.removeFirst()$ 
%\State $\vec{x}\gets\vec{o}$ 
\If {$ut > maxVal$ \textbf{or} $(ut = maxVal$ \textbf{and} $u_a(\vec{y})>u_a(\vec{o}))$}
        %\State $L.add(\{\vec{x},\vec{s},ut\})$	
    		\State $maxVal \gets ut$ 
    		\State $\vec{o}\gets\vec{y}$
    		\State $L.prune(maxVal)$ 
\EndIf

\ForAll{$i \in \text{conflicts}(\vec{x})$} %\mathcal{C}$}\Comment{For all the remaining conflicts}
%\State $prod\gets u_a(\vec{x})\times u_b(\vec{x})$
%\If {$x_i = *$}
	\For{$action \in \{0,1\}$}
	\State $x[i]\gets action$
    \State $\{ut,\vec{y}\}=greedySolution(\vec{x})$ 
    \If {$ut > maxVal$ \textbf{or} $(ut = maxVal$ \textbf{and} $u_a(\vec{y})>u_a(\vec{o}))$}  	
        \State $L.add(\{\vec{x},\vec{y},ut\})$	
    		%\State $maxVal \gets prod$ 
    		%\State $maxTarget\gets i$
    		%\State $maxAction\gets x[i]$
	\EndIf
	\EndFor
	\State $x[i]\gets *$
%\EndIf
\EndFor
%\State $o[maxTarget]=maxAction$ %\Comment{The conflict that causes the least utility loss is solved}
%\State $\mathcal{C}\gets\mathcal{C}\setminus \{maxTarget\}$
\EndWhile
\end{algorithmic}
\end{footnotesize}
\caption{GreedyBnB Heuristic - Agent a}
\label{al:pg3}
\end{algorithm}

%the greedy algorithm can be used as a selection algorithm to prioritize options within a search, or branch and bound algorithm  

Algorithm \ref{al:pg3} lists our GreedyBnB heuristic. Firstly, it constructs a partial action vector $\vec{t}$ with all the conflicts detected (Lines 1-7). Secondly, it uses the greedy heuristic to obtain a solution to the partial action vector created (Line 9). Then, it adds the partial action vector, the greedy solution and its utility to the list $L$ (Line 10), which is ordered by decreasing values of utility of the greedy solution obtained. While $L$ is not empty, we retrieve the first element of the ordered list. If this has higher utility than the best solution seen so far, the solution is recorded as the best one seen so far and all the nodes in the list that have less utility are pruned (Lines 14-18). After this, we generate all the possible partial action vectors that arise from considering the possible actions (granting/1 or denying/0) for all the remaining conflicts in the partial action vector retrieved from the list (Lines 19-28). If the greedy solution for any of these partial action vectors produces a higher utility than the best solution seen so far, we add this partial action vector (which will have one conflict less than the partial action vector retrieved from the list), its greedy solution and the utility of the solution to the ordered list (Lines 22-25). 

%If there is one partial action vector with utility 

%Informally speaking, we try to find branches different from the one originally chosen by the greedy heuristic that could lead to better outcomes. 

Finally, it is worth noting that the GreedyBnB heuristic can also be implemented in a similar way to anytime algorithms. These algorithms can be interrupted before they end but will still provide a valid solution if interrupted, and are expected to produce increasingly good solutions the more time they are given to run. In particular, we could implement the heuristic so that it stops after a given amount of time, and this is explored further below in Section \ref{sec:results}. When the heuristic stops, it will return the best solution seen so far. In the experiments section we describe the performance achieved with the heuristic at various cutoff points.

\section{Experimental Results}
\label{sec:results}
%In this section, we detail the experiments that 
%We conducted a series of experiments to empirically compare the performance of our proposed negotiation mechanism without the heuristic to its use with the heuristic, contrasting the effects of different \emph{importance threshold} $\varphi$ values within the heuristic. We assessed the performance of both approaches with respect to the number of action vectors considered to find a solution, the percentage of agreements achieved, and the maximum product of utilities obtained.  

We conducted a series of experiments to compare the performance of our proposed negotiation mechanism with and without heuristics empirically by measuring the number of action vectors considered and the execution time required to find a solution, as well as the maximum product of utilities obtained. 

\subsection{Experimental Setting}
We implemented our mechanism and heuristics in Java and report experiments conducted on a 3.1 GHz Intel Core i5 iMac with 8 GB RAM. The rationale for the parameters explained below was the following: wherever we knew the real value or distribution of one of the parameters in real Social Media we used that to inform the values considered; and if we did not know the real value of distribution of the parameters considered then we used a different random value for each experiment conducted. For the possible set of parameters, we performed 1000 different experiments in order to support the findings with statistically significant evidence. 

We considered 3 relationship types in line with related literature on online communities and social networks \cite{raad2013discovering,hogg2008multiple}. Besides, we considered different numbers of target agents --- 10 to 200 in increments of 10 --- based on typical real-world values, given that the average user on Facebook has 130 friends \cite{quercia2012personality}. We also considered a maximum intimacy value $\mathcal{Y}$ of 10, i.e.\ an intimacy range of [0,10]. Note that the maximum intimacy value does not have any effect on the performance of the mechanism, as it just defines a range of real values for the possible intimacy values. If another range of real values is used, the intimacy values would be scaled but the results would be the same in terms of the solution agreed. %For every number of target agents, we performed 1000 different experiments in order to support the findings with statistically significant evidence.

For the parameters we did not have the real data, we just randomised them. That is, we generated a random matrix of intimacies among agents, random assignment of agents to relationship types, and random privacy policies for both negotiating agents. Moreover, we ensured that for each situation, there was at least one conflict --- if not, we reinitialised the matrix and privacy policies until at least one conflict occurred. 

After this, we ran the negotiation mechanism with and without heuristics to obtain the solution to this situation and accounted for the number of action vectors each approach needed to explore and, where available, the utility of the solution achieved and the execution time needed to achieve the solution. As explained below, there were situations in which not all the approaches were able to produce an actual solution in reasonable time. 

Finally, when using the mechanism with heuristics, we varied the parameters of the particular heuristic should they have one. The heuristics that can be parametrised are the distance-based heuristic and the GreedyBnB heuristic. For the distance-based heuristic one can use a different importance threshold $\varphi$ as described in Section 7.1. In particular, we show the performance of different values for this parameter in Sections 8.2, 8.3, and 8.4. This values were $\varphi=\{0.5,1, 2, 3, 4\}$. Higher values for the importance threshold did not remove enough action vectors and took approximately the same time as the mechanism without heuristics, while lower values for the importance threshold had too much utility loss. Regarding the GreedyBnB heuristic, it can be interrupted before it ends but will still provide a valid solution if interrupted as described in Section 7.3 --- though it is expected to produce increasingly good solutions the more time they are given to run. In section 8.5, we illustrate the performance achieved with the heuristic at various cutoff points (after 30, 50, 100, 200, and 500 ms). More cutoff points were considered, but these were the most representative ones to be shown in the figures.

\subsection{Number of Action Vectors}
\begin{figure}[h]
\centering
%\hspace{-10pt}
\vspace{-20pt}

\includegraphics[scale=0.33]{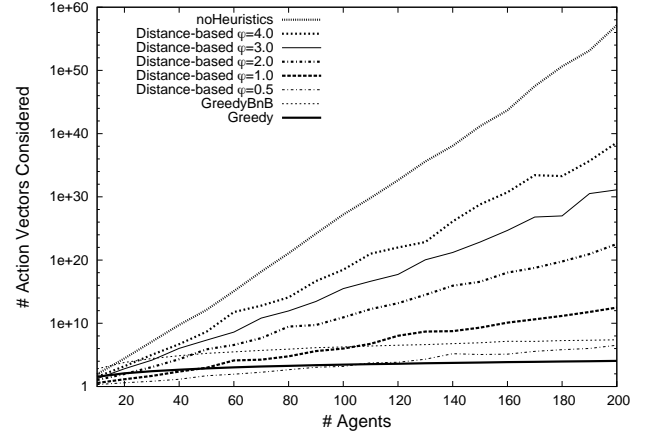}
\label{fig-np}
\vspace{-35pt}

\caption{Average number of action vectors considered per number of agents. Note that Y-axis uses a logarithmic scale. %, because in a normal scale we could have only appreciated the number of action vectors when using the %mechanism without the heuristic. 
%We only show results for one relationship type, because the number of action vectors considered is the same regardless of the number of relationship types (they are only used to calculate total utility).
}
\end{figure}

We measured the number of action vectors to come up with a reasonable estimate of expected execution time, because for more than 40 target agents the experiments where heuristics were not used would not finish after 7 days running. This indicated that without heuristic pruning the negotiation mechanism is not usable in practice. Figure 2 shows the average number of action vectors (including partial action vectors when using the heuristics) that each approach would need to consider to solve the problem given a number of agents. %We measured the number of action vectors to come up with a reasonable estimate of expected execution time, because for more than 40 target agents the experiments where heuristics were not used would not finish after 7 days running. This indicated that without heuristic pruning the negotiation mechanism is not usable in practice. 
The plot also shows that the lower $\varphi$ for the distance-based heuristic, the lower the number of action vectors considered. This is because for low $\varphi$ values more actions are detected as being more important for one negotiating agent than for the other, so fewer action vectors are generated. We can also observe that both Greedy and GreedyBnB perform similarly, though Greedy considers fewer action vectors. Obviously, GreedyBnB requires more action vectors because it needs to invoke the Greedy heuristics several times to compute a solution. 

%both the Greedy and the GreedyBnB heuristics perform very similar. However, GreedyBnB requires more action vectors because, obviously, it needs to run several times the Greedy heuristics for each experiment. 

%as it can be seen, in some cases it would take too much time to run the actual experiments due to the exponential nature of the problem --- indeed, for the utility product results in Figures 2 and 3, we could only show the results obtained in experiments that lasted at most 5 days. Clearly, without heuristic pruning the negotiation mechanism is not usable in practice for more than around 40 target agents --- it would involve considering an average of more than $10^{10}$ action vectors. Note that we only depict the results for 1 relationship type, because the number of action vectors is the same regardless the number of relationship types (they are only used to reckon utility).  %Moreover, the lower $\varphi$, the lower the number of action vectors considered. This is because for low $\varphi$ values more actions are detected as being more important for one negotiating agent than for the other, so less action vectors are generated.

\subsection{Utility}

\begin{figure}[h]
%\left
%\hspace{-10pt}
\centering
%\hspace{-10pt}
\vspace{-20pt}

\includegraphics[scale=0.33]{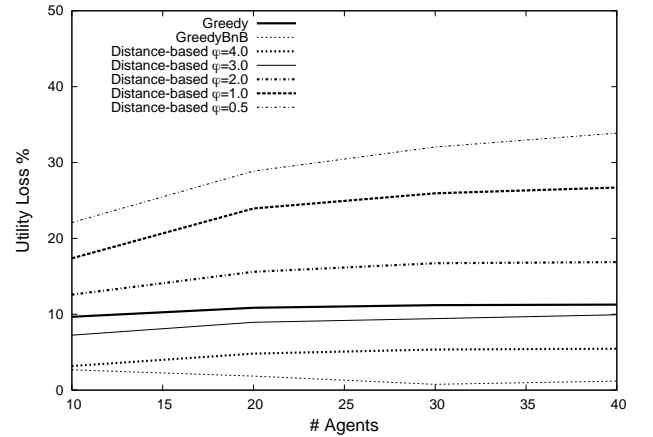}
\label{fig:loss}
\vspace{-35pt}

\caption{Average \% of utility (product) loss for different numbers of target agents. As before, we were only able to obtain results for up to 40 target agents without heuristics.}
\end{figure}

\begin{figure}[!h]
\centering
%\hspace{-10pt}
\vspace{-20pt}

\includegraphics[scale=0.33]{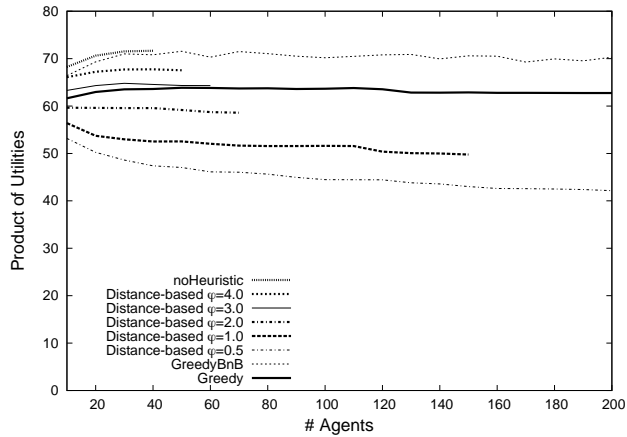}
\label{fig:ut}
\vspace{-35pt}

\caption{Utility product for different numbers of target agents, heuristic and non-heuristic approaches.}
\end{figure}

Figure 3 shows the percentage of utility lost when using heuristics compared to the optimal solution. All heuristics have an impact on optimality, but the best results are obtained with the GreedyBnB heuristic and the distance-based heuristic with $\varphi=4$. However, if we consider the number of action vectors needed considered before returning a solution (as shown in Figure 2), the GreedyBnB heuristic is the one that generally offers the best tradeoff between number of action vectors and utility loss. Moreover, in order to determine how well the heuristics scale when increasing the number of target agents, Figure 4 shows the total product of utilities achieved by each approach. For $\varphi\in\{1,2,3,4\}$ we had the same problem as without use of any heuristic, i.e., we could not obtain results in reasonable time for more than a small number of target agents. In contrast to this, the Greedy and GreedyBnB heuristics scale by far better and clearly outperform the distance-based heuristic with $\varphi=0.5$, which prunes enough action vectors to produce results in reasonable time for all numbers of target agents considered. We can also observe that when the number of targets increases, GreedyBnB loses less percent of utility with respect to the optimal one. Finally, we can also observe that GreedyBnB clearly outperforms Greedy in terms of utility. Greedy algorithms mostly (but not always) fail to find the globally optimal solution, because they usually do not  exhaustively consider the whole space, and may make commitments to certain choices too early, which prevent them from finding the best overall solution later. Using GreedyBnB, we consider branches of the search space that were initially discarded by Greedy which usually leads to finding a better solution. However, exploring more branches of the search space also comes at a cost --- as we have seen previously, GreedyBnB obviously needs to consider many more action vectors than Greedy. In the following section, we assess the difference between these two in terms of actual execution time.

\begin{figure}[!h]
\centering
%\hspace{-10pt}
\vspace{-20pt}

\includegraphics[scale=0.33]{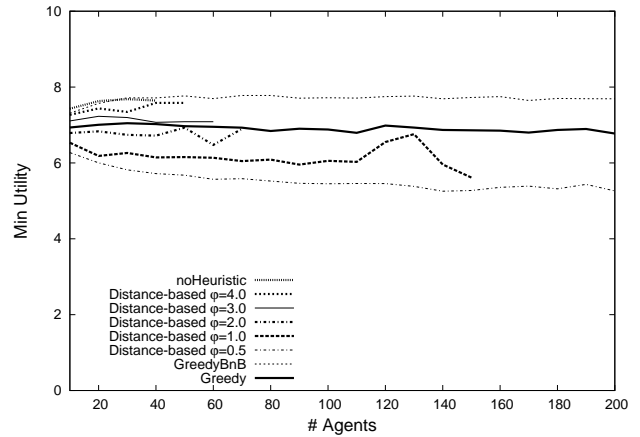}
\label{fig:ut}
\vspace{-35pt}

\caption{Minimum utility for different numbers of target agents, heuristic and non-heuristic approaches.}
\end{figure}

Finally, we sought to assess the quality of the results obtained in terms of to what extent individual privacy preferences were being covered. To this aim, we also considered what is the minimum utility achieved among negotiating agents for each deal, i.e., which is the utility for the negotiating agent that is least \emph{favoured} in the negotiation. Figure 5 shows the minimum utility achieved among negotiating agents for each deal considering different numbers of target agents, and heuristic and non-heuristic approaches. We can see that, without heuristics, the negotiating agent that ends up with the lowest utility will always have a utility of at least 7.5. Considering the utility function (Definition 8) and the maximum intimacy value $\mathcal{Y}$ of 10, the maximum utility that negotiating agents can achieve is 10, though this would only be for the case that the deal chosen is the one that favours them the most, which is not possible as compromises are usually done to achieve an agreement. Therefore, without heuristics the \emph{least favoured} negotiating agent has always at least 75\% of their preferences covered --- in other words, it will only lose at most 25\% in utility. Moreover, this increases to $\approx$ 78\% as the number of target agents increases, because there are more opportunities to find better compromises. Regarding the use of the mechanism with heuristics, the GreedyBnB is again the best heuristic, so that the mechanism with that heuristic provides roughly the same values as the mechanism without heuristics.

\subsection{Execution Time}
As we were able to obtain results in reasonable time for all the possible configurations when Greedy and GreedyBnB heuristics were used, we were able to compare them in terms of execution time to complement the results obtained in terms of number of action vectors needed to compute a solution. We obviated the distance-based heuristic here because we were only able to obtain results in reasonable time with $\varphi=0.5$, but with this $\varphi$ value, the heuristic produced results that are very far from the optimal (up to more than 30\%).

\begin{figure}[!h]
\label{fig:time}
\centering
%\hspace{-10pt}
\vspace{-20pt}

\includegraphics[scale=0.33]{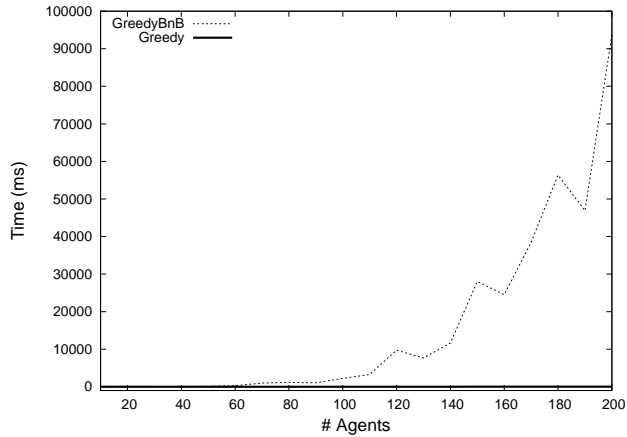}
\vspace{-35pt}

\caption{Execution time for Greedy and GreedyBnB heuristics.}
\end{figure}

Figure 4 shows execution times for the Greedy and GreedyBnB heuristics. We can observe that, from 120 target agents upwards, the GreedyBnB heuristic takes more than $\approx 100$ seconds to compute a solution while the Greedy heuristic needs less than $30ms$. Clearly, adding 100 seconds to the process of posting an item could be considers too much for users of social media. However, the Greedy heuristic, which would be fast enough, usually implies a loss of utility of around 10\%. Thus, use of any of these two heuristics might be ultimately unsuitable for implementation in a practical tool to resolve actual privacy policy conflicts in social media. Can we find a method that lies somewhere ``between'' these two that better balances the tradeoff between execution time and optimality?

\subsection{Exploiting GreedyBnB anytime capabilities}
To come up with such a method,  we repeated the experiments, this time exploiting the anytime capabilities of GreedyBnB. In particular, we limited the available computation time to a specific bound as suggested in section \ref{sec:heuristic}, i.e., we stopped looking for further possible solutions after a given amount of time elapsed. 

%\subsection{Execution Time}
\begin{figure}[!h]
\label{fig:anytime}
\centering
%\hspace{-10pt}
\vspace{-20pt}

\includegraphics[scale=0.33]{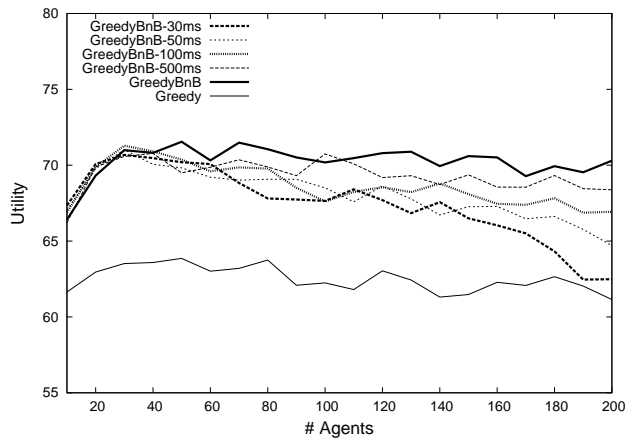}
\vspace{-35pt}

\caption{Utility considering different time bounds for GreedyBnB heuristic.}
\end{figure}

Figure 5 shows the utility achieved for some temporal bounds compared to that obtained using the unbounded version of GreedyBnB. It also shows the results for the Greedy heuristic as a lower bound for the utility achieved. As expected, the more time GreedyBnB is given, the better the results obtained. Importantly, we can see that with 500ms the results obtained would be very similar to unbounded GreedyBnB, and much better than those achieved using the Greedy heuristic. For instance, for 200 agents the difference between running unbounded  GreedyBnB (with a runtime of 100 seconds) and its bounded version limited to 500ms is less than 3\%. 

%-Greedy would be always the lower bound with approximately 10\% of utility loss with respect to the optimal. quan mes temps li donem al anytime pues mes acurada sera, pero es pot vore com per a 200 agents, 500ms traurien una solucio similar a la obtesa amb 100000ms  

\subsection{Discussion}
The main conclusion from our experiments is that the negotiation mechanism cannot be used in practice for realistic numbers of target agents without the complexity reduction provided by our heuristics. For 200 target agents, for example, we would need to consider $\approx 10^{50}$ action vectors. Moreover, although for a limited number of target agents the distance-based heuristic performs well for some parameter choices, the greedy-based heuristics generally outperforms it both in terms of complexity reduction and in terms of loss of utility, at least for larger numbers of agents. In particular, using the greedy heuristic seems to achieve very good results regarding the number of action vectors considered and execution time --- for 200 agents it would consider only $\approx 10^3$ action vectors and would take 30ms, for example --- while incurring a loss of optimality that could acceptable: the heuristics would allow us to reach agreement with 200 target agents while sacrificing only around 10\% of utility compared to the non-heuristic version of the algorithm. Using the GreedyBnB heuristic achieves outstanding results in terms of loss of optimality but is worse than Greedy in terms of the number of action vectors considered and execution time required. However, choosing the appropriate parameters to bound the execution time of GreedyBnB seems to achieve the best tradeoff between optimality and execution time. For instance, for 200 agents the bounded version of GreedyBnB would sacrifice only around 3\% of the utility obtained while taking only 500ms, which seems to be fast enough for actual use in common social media applications. 

%We consider this an acceptable tradeoff in exchange for the ability to solve the problem in reasonable time --- for 200 agents it takes less than 1 second (an average of 30ms) to achieve agreement using the greedy heuristics. 

%To summarise, the negotiation mechanism cannot be used in practice for realistic numbers of target agents without the complexity reduction provided by our heuristics. For 200 target agents, for example, we would need to consider $\approx 10^{50}$ action vectors. Moreover, although for a limited number of target agents the distance-based heuristic performs well for some parameter choices, the greedy-based heuristics generally outperform the distance-based heuristic both in complexity reduction and in loss of utility for larger numbers of agents. In particular, using the greedy heuristic seems to achieve good results regarding the number of action vectors considered --- for 200 agents it would consider only $\approx 190$ action vectors, for example --- while incurring a loss of optimality that is acceptable: the heuristics would allow us to reach agreement with 200 target agents while only sacrificing around 10\% of utility as compared to the non-heuristic version of the algorithm. We consider this an acceptable tradeoff in exchange for the ability to solve the problem in reasonable time --- for 200 agents it takes less than 1 second (an average of 30ms) to achieve agreement using the greedy heuristics. 

\section{Related Work}
\label{sec:relwork}
%Until now, very few contributions have considered the problem of detecting and resolving conflicts in privacy policies for shared items between users of a SNS. %In the following we detail how these proposals differ from ours. 
Over the last few years, many studies have been devoted to improving user privacy in social media. Until now, these mechanisms have often been shown not to effectively protect privacy and sensitive information \cite{zheleva09, madejski11}. To address this problem, many approaches have recently emerged, e.g.\ \cite{carminati2009enforcing, fong2011relationship,fang2010privacy,ali2014self}. In particular, AI and Multi-agent Systems approaches have been suggested as appropriate to control socio-technical systems like social media \cite{singh2013norms} and as a foundation for social computing%\footnote{Note that \cite{rovatsos2014multiagent} does not share any commonality with this paper.} 
\cite{rovatsos2014multiagent}. Such et al. \cite{such13review} present an extensive review of agent and Multi-agent Systems approaches that deal with preserving privacy%\footnote{Note that \cite{such13review} is a review paper, which does not propose any new model and does not share any commonality with this paper.}
. An example of these kind of approaches to manage privacy in Social Media is the work of Krupa and Vercouter \cite{krupa2012handling,ciortea2012designing}, which proposes an agent-based framework to control information flows in Social Networks. Another example is the work of K{\"o}kciyan and Yolum \cite{kokciyancommitment}, which proposed an agent-based framework for privacy management and detection of violations based on commitments. Finally, Such et al. \cite{such12sddm} proposed the first agent-based mechanism to decide whether personal information is shared and with whom based on information-theoretic measures of intimacy and privacy
%\footnote{Note that paper \cite{such12sddm} does not share any common parts with this paper, nor the model presented in \cite{such12sddm} resembles in any way to the mechanism proposed in this paper.}
. However, all of these approaches do not consider the problem of items that may affect more than one user, and do not support multiple users agreeing on to whom these items are shared. Therefore, these approaches do not consider the privacy preferences of all of the users affected by an item when deciding to whom and whether or not information is shared.

Few works have actually been proposed to deal with the problem of collaboratively defining  privacy policies for shared items between two or more users of a social media site. We shall discuss them and how they relate to our work in the following paragraphs. %now discuss them. %One of the most illustrating examples is that of two users that define a common privacy policy for a photo in which both users are depicted.  

Wishart et al.\ \cite{wishart2010collaborative} propose a method to define privacy policies collaboratively. Their approach is based on a collaborative definition of privacy policies in which all of the parties involved can define strong and weak preferences. They define a privacy language to specify users' preferences in the form of strong and weak conditions, and they detect privacy conflicts based on them. However, this approach does not involve any automated method to resolve conflicts, only some suggestions that users may want to consider consider when they try to resolve such conflicts manually.  
%In \cite{wishart2010collaborative} the authors define a privacy language to specify users' preferences in the form of strong and weak conditions, and they detect privacy conflicts based on them. However, this approach does not involve any automated method to solve conflicts, only some suggestions that the users could consider when they try to resolve conflicts manually.

The work described in \cite{squicciarini2009collective} %presents an approach to collaborative privacy policy authoring in social networks. It 
is based on an incentive mechanism where users are rewarded with a quantity of numeraire each time they share information or acknowledge other users (called co-owners) who are affected by the same item% (e.g.\ a photo in which all co-owners appear)
. When there are conflicts %regarding the privacy policy to be applied 
among co-owners' policies, the use of the Clark Tax mechanism is suggested, where users can spend their numeraire bidding for the policy that is best for them. As stated in \cite{hu2012detecting}, the usability of this approach may be limited, because users could have difficulties in comprehending the  mechanism and specify appropriate bid values in auctions. Moreover, the auction process adopted in their approach implies that only the winning bid determines who will be able to access the data, instead of accommodating all stakeholders' privacy preferences. %Moreover, as the numeraire is not \emph{real} money in this domain,  they could have the problem that people that won much numeraire in the past(because they shared much information) will have more numeraire to spend it at will, possibly leading to imposing their desired policies to others. Moreover, people who share more information are usually less concerned about their privacy. Therefore, it could turn out that the less concerned are the most powerful in the negotiation process, which is clearly detrimental to more concerned users. %Instead, we propose an approach in which 

In \cite{hu2012detecting}, %the authors define a collective mechanism for authoring privacy policies. The 
users must manually define their \emph{trust} to other users, the sensitivity that each of the items has for them, and their general privacy concern. Then, the authors use these parameters to calculate two main measures, privacy risk and sharing loss. In particular, they calculate the privacy risk and the sharing loss on what they call segments --- in our terminology, a segment equals the set of agents in conflict --- as a whole, i.e.\ all of the agents in these segments are assigned the action preferred by either one party or the other in the negotiation. That is, in our terminology only two action vectors --- $\vec{v}$ and $\vec{w}$ induced by the privacy policies $P_a$ and $P_b$ respectively --- are considered, and the action vector chosen is the one that maximises the tradeoff between privacy risk and sharing loss. Clearly, not considering other possible action vectors could lead to outcomes that are far from optimal. 

Finally, there are also related approaches based on voting in the literature \cite{carminati2011collaborative,thomas2010unfriendly,hu2012multiparty}. In these cases, a third party collects the decision to be taken (granting/denying) for a particular friend from each party. Then, the authors propose to aggregate a final decision based on one voting rule (majority, veto, etc.). However, the rule to be applied is either fixed \cite{carminati2011collaborative,thomas2010unfriendly} or is chosen by the user that uploads the item \cite{hu2012multiparty}. The problem with this is that the solution to the conflicts then becomes a unilateral decision (being taken by a third-party or by the user that uploads the item) and, thus, there is no room for users to actually negotiate and achieve compromise themselves.  %already been described in Section \ref{sec:evaluation} --- i.e., uploader overwrites (UO), majority voting (MV), and veto voting (VV). These approaches are static, in the sense that they always aggregate individual votes in the same way by following the same voting rule. Thus, these approaches are unable to adapt to different situations that can motivate different concessions by the negotiating users, which makes these approaches unable to suggest acceptable results most of the times, as shown in Section 7. Only in \cite{}, the authors consider that, after doing some other calculations, a different voting rule could be applied depending on the situation. However, it is the user who uploads/posts the item the one who chooses \emph{manually} which one of the voting rules (UO,MV,VV) to apply for each item. %This could be seen as a basic, though rudimentary, adaptation. 
%The main problem with this --- apart from having to specify the voting rule manually for every item --- is that the choice of the voting rule to be applied is unilateral. That is, the user that uploads the item decides the rule to apply without considering the rest of the negotiating users' preferences, which becomes a unilateral decision on a multi-party setting. %Hence, chances are that the voting rule chosen by the user that uploads the item is not the most appropriate for the specific situation. Indeed, 
Moreover, in the latter case, it might actually be quite difficult for the user that uploads the item to anticipate which voting rule would produce the best result without knowing the preferences of the other users.

\section{Conclusions}
\label{sec:conclusions}
We presented an automated method for detecting and resolving privacy policy conflicts in social media applications. To resolve conflicts, we proposed the use of an automated negotiation mechanism. This mechanism is based on the intimacy among agents, which determines the utility of other agents' proposals. In using intimacy as the determining factor for utility, we followed the findings of most empirical studies with real users, which confirm that it is the main factor that influences their behaviour with regard to setting privacy policies. This suggests that our approach is sound in terms of modelling real users' preferences. 

Moreover, in order to reduce the complexity of the negotiation mechanism proposed, we proposed three heuristics and showed through an experimental evaluation comparing the performance of the negotiation mechanism proposed with and without heuristics that: (i) use of the negotiation mechanism is not practicable without heuristics; (ii) the distance-based heuristic only performs well for a limited number of target agents; (iii) the greedy heuristic offers good tradeoffs between complexity and optimality when scaling up the number of target agents; and (iv) the best heuristic overall is GreedyBnB with a time bound. In particular, GreedyBnB bounded to 500ms would sacrifice only 3\% in optimality in our experiments. %We are currently working on generalising our mechanism for more than two negotiating agents. %, because shared items may concern more than two users. %while inducing only negligible incompleteness effects, provided the appropriate parameters are chosen. 
%- Incloure Anytime GreedyHeuristic Aci igual si conclous respecte a esta i au millor.

%We are currently working on implementing our proposed conflict detection and resolution method as a Facebook application, combining it with other existing tools that are able to automatically elicit intimacy and relationship types \cite{fogues2013bff}. This Facebook application will also be used to evaluate human acceptability of the solutions proposed by our negotiation mechanism. 
The research presented in this article is a stepping stone towards automated privacy policy negotiation. 
An interesting future line of research would be to consider the fact that disclosing items can also make relationships evolve \cite{such12sddm}, which could play a role in shaping users' preferences about disclosure and negotiation outcomes. Finally, we would also like to extend our mechanism in order to consider the intimacy between the negotiating parties, which is known to influence negotiation strategies and, in particular, may determine to what extent negotiating parties are willing to concede during a negotiation \cite{sierra07}.

\bibliographystyle{IEEEtran}
\bibliography{IEEEabrv,policy-negotiation}

\begin{thebibliography}{10}
\providecommand{\url}[1]{#1}
\csname url@samestyle\endcsname
\providecommand{\newblock}{\relax}
\providecommand{\bibinfo}[2]{#2}
\providecommand{\BIBentrySTDinterwordspacing}{\spaceskip=0pt\relax}
\providecommand{\BIBentryALTinterwordstretchfactor}{4}
\providecommand{\BIBentryALTinterwordspacing}{\spaceskip=\fontdimen2\font plus
\BIBentryALTinterwordstretchfactor\fontdimen3\font minus
  \fontdimen4\font\relax}
\providecommand{\BIBforeignlanguage}[2]{{%
\expandafter\ifx\csname l@#1\endcsname\relax
\typeout{** WARNING: IEEEtran.bst: No hyphenation pattern has been}%
\typeout{** loaded for the language `#1'. Using the pattern for}%
\typeout{** the default language instead.}%
\else
\language=\csname l@#1\endcsname
\fi
#2}}
\providecommand{\BIBdecl}{\relax}
\BIBdecl

\bibitem{gross2005information}
R.~Gross and A.~Acquisti, ``Information revelation and privacy in online social
  networks,'' in \emph{Proceedings of the 2005 ACM workshop on Privacy in the
  electronic society ({WPES})}.\hskip 1em plus 0.5em minus 0.4em\relax ACM,
  2005, pp. 71--80.

\bibitem{stutzman2013silent}
F.~Stutzman, R.~Gross, and A.~Acquisti, ``Silent listeners: The evolution of
  privacy and disclosure on facebook,'' \emph{Journal of Privacy and
  Confidentiality}, vol.~4, no.~2, p.~2, 2013.

\bibitem{bonneau2010privacy}
J.~Bonneau and S.~Preibusch, ``The privacy jungle: On the market for data
  protection in social networks,'' in \emph{Economics of information security
  and privacy}.\hskip 1em plus 0.5em minus 0.4em\relax Springer, 2010, pp.
  121--167.

\bibitem{quercia2012personality}
D.~Quercia, R.~Lambiotte, M.~Kosinski, D.~Stillwell, and J.~Crowcroft, ``The
  personality of popular facebook users,'' in \emph{Proceedings of the ACM 2012
  conference on Computer Supported Cooperative Work (CSCW'12)}, 2012, pp.
  955--964.

\bibitem{staddon2012privacy}
J.~Staddon, D.~Huffaker, L.~Brown, and A.~Sedley, ``Are privacy concerns a
  turn-off?: engagement and privacy in social networks,'' in \emph{Proceedings
  of the Eighth Symposium on Usable Privacy and Security ({SOUPS})}.\hskip 1em
  plus 0.5em minus 0.4em\relax ACM, 2012, pp. 1--13.

\bibitem{carminati2009enforcing}
B.~Carminati, E.~Ferrari, and A.~Perego, ``Enforcing access control in
  web-based social networks,'' \emph{ACM Transactions on Information and System
  Security (TISSEC)}, vol.~13, no.~1, p.~6, 2009.

\bibitem{fong2011relationship}
P.~W. Fong, ``Relationship-based access control: protection model and policy
  language,'' in \emph{Proceedings of the first ACM conference on Data and
  application security and privacy ({CODASPY})}.\hskip 1em plus 0.5em minus
  0.4em\relax ACM, 2011, pp. 191--202.

\bibitem{wang2012isac}
Y.~Wang, E.~Zhai, E.~K. Lua, J.~Hu, and Z.~Chen, ``isac: Intimacy based access
  control for social network sites,'' in \emph{Proceedings of the 2012 9th
  International Conference on Ubiquitous Intelligence \& Computing and 9th
  International Conference on Autonomic \& Trusted Computing (UIC/ATC)}.\hskip
  1em plus 0.5em minus 0.4em\relax IEEE, 2012, pp. 517--524.

\bibitem{houghton2010privacy}
D.~J. Houghton and A.~N. Joinson, ``Privacy, social network sites, and social
  relations,'' \emph{Journal of Technology in Human Services}, vol.~28, no.
  1-2, pp. 74--94, 2010.

\bibitem{wiese2011you}
J.~Wiese, P.~G. Kelley, L.~F. Cranor, L.~Dabbish, J.~I. Hong, and J.~Zimmerman,
  ``Are you close with me? are you nearby? investigating social groups,
  closeness, and willingness to share,'' in \emph{Proceedings of the 13th
  international conference on Ubiquitous computing ({UbiComp})}.\hskip 1em plus
  0.5em minus 0.4em\relax ACM, 2011, pp. 197--206.

\bibitem{duck2007human}
S.~Duck, \emph{Human relationships}.\hskip 1em plus 0.5em minus 0.4em\relax
  SAGE Publications Limited, 2007.

\bibitem{strahilevitz2005social}
L.~J. Strahilevitz, ``A social networks theory of privacy,'' in \emph{American
  Law \& Economics Association Annual Meetings}.\hskip 1em plus 0.5em minus
  0.4em\relax bepress, 2005, p.~42.

\bibitem{gates2007access}
C.~Gates, ``Access control requirements for web 2.0 security and privacy,''
  \emph{IEEE Web}, vol.~2, no.~0, 2007.

\bibitem{such13review}
J.~M. Such, A.~Espinosa, and A.~Garc\'{i}a-Fornes, ``A survey of privacy in
  multi-agent systems,'' \emph{Knowledge Engineering Review}, vol.~29, pp.
  313--344, 2014.

\bibitem{rosenschein94}
\BIBentryALTinterwordspacing
J.~S. Rosenschein and G.~Zlotkin, \emph{Rules of encounter: designing
  conventions for automated negotiation among computers}, ser. Artificial
  intelligence.\hskip 1em plus 0.5em minus 0.4em\relax MIT Press, 1994.
  [Online]. Available: \url{http://books.google.com/books?id=4ZhPMk3ftpoC}
\BIBentrySTDinterwordspacing

\bibitem{fang2010privacy}
L.~Fang and K.~LeFevre, ``Privacy wizards for social networking sites,'' in
  \emph{Proceedings of the 19th international conference on World Wide Web
  (WWW)}.\hskip 1em plus 0.5em minus 0.4em\relax ACM, 2010, pp. 351--360.

\bibitem{squicciarini2011a3p}
A.~C. Squicciarini, S.~Sundareswaran, D.~Lin, and J.~Wede, ``A3p: adaptive
  policy prediction for shared images over popular content sharing sites,'' in
  \emph{Proceedings of the 22nd ACM conference on Hypertext and hypermedia
  ({H}ypertext)}, 2011, pp. 261--270.

\bibitem{danezis2009inferring}
G.~Danezis, ``Inferring privacy policies for social networking services,'' in
  \emph{Proceedings of the 2nd ACM workshop on Security and artificial
  intelligence}.\hskip 1em plus 0.5em minus 0.4em\relax ACM, 2009, pp. 5--10.

\bibitem{granovetter1973strength}
M.~Granovetter, ``The strength of weak ties,'' \emph{American journal of
  sociology}, vol.~78, no.~6, pp. 1360--1380, 1973.

\bibitem{green06}
K.~Green, V.~J. Derlega, and A.~Mathews, ``Self-disclosure in personal
  relationships,'' in \emph{The Cambridge Handbook of Personal
  Relationships}.\hskip 1em plus 0.5em minus 0.4em\relax Cambridge University
  Press, 2006, pp. 409--427.

\bibitem{white2002navigability}
D.~R. White and M.~Houseman, ``The navigability of strong ties: Small worlds,
  tie strength, and network topology,'' \emph{Complexity}, vol.~8, no.~1, pp.
  72--81, 2002.

\bibitem{gilbert2009predicting}
E.~Gilbert and K.~Karahalios, ``Predicting tie strength with social media,'' in
  \emph{Proceedings of the 27th international conference on Human factors in
  computing systems (CHI)}.\hskip 1em plus 0.5em minus 0.4em\relax ACM, 2009,
  pp. 211--220.

\bibitem{fogues2013bff}
\BIBentryALTinterwordspacing
R.~L. Fogu{\'e}s, J.~M. Such, A.~Espinosa, and A.~Garcia-Fornes,
  ``\BIBforeignlanguage{English}{Bff: A tool for eliciting tie strength and
  user communities in social networking services},''
  \emph{\BIBforeignlanguage{English}{Information Systems Frontiers}}, vol.~16,
  no.~2, pp. 225--237, 2014. [Online]. Available:
  \url{http://dx.doi.org/10.1007/s10796-013-9453-6}
\BIBentrySTDinterwordspacing

\bibitem{gilbert2012predicting}
\BIBentryALTinterwordspacing
E.~Gilbert, ``Predicting tie strength in a new medium,'' in \emph{Proceedings
  of the ACM 2012 conference on Computer Supported Cooperative Work}, ser. CSCW
  '12.\hskip 1em plus 0.5em minus 0.4em\relax New York, NY, USA: ACM, 2012, pp.
  1047--1056. [Online]. Available:
  \url{http://doi.acm.org/10.1145/2145204.2145360}
\BIBentrySTDinterwordspacing

\bibitem{cranor2005security}
L.~Cranor and S.~Garfinkel, \emph{Security and usability: designing secure
  systems that people can use}.\hskip 1em plus 0.5em minus 0.4em\relax O'Reilly
  Media, Incorporated, 2005.

\bibitem{lopes2008negotiation}
F.~Lopes, M.~Wooldridge, and A.~Q. Novais, ``Negotiation among autonomous
  computational agents: principles, analysis and challenges,'' \emph{Artificial
  Intelligence Review}, vol.~29, no.~1, pp. 1--44, 2008.

\bibitem{vidal2010}
J.~M. Vidal, \emph{Fundamentals of multiagent systems with NetLogo examples.},
  2010.

\bibitem{osborne1990bargaining}
M.~J. Osborne and A.~Rubinstein, \emph{Bargaining and markets}.\hskip 1em plus
  0.5em minus 0.4em\relax Academic press San Diego, 1990, vol.~34.

\bibitem{endriss2006monotonic}
U.~Endriss, ``Monotonic concession protocols for multilateral negotiation,'' in
  \emph{AAMAS}.\hskip 1em plus 0.5em minus 0.4em\relax ACM, 2006, pp. 392--399.

\bibitem{raad2013discovering}
E.~Raad, R.~Chbeir, and A.~Dipanda, ``Discovering relationship types between
  users using profiles and shared photos in a social network,''
  \emph{Multimedia Tools and Applications}, vol.~64, no.~1, pp. 141--170, 2013.

\bibitem{hogg2008multiple}
T.~Hogg, D.~M. Wilkinson, G.~Szabo, and M.~J. Brzozowski, ``Multiple
  relationship types in online communities and social networks.'' in \emph{AAAI
  Spring Symposium: Social Information Processing}, 2008, pp. 30--35.

\bibitem{zheleva09}
E.~Zheleva and L.~Getoor, ``To join or not to join: the illusion of privacy in
  social networks with mixed public and private user profiles,'' in
  \emph{Proceedings of the 18th international conference on World wide web
  {(WWW)}}.\hskip 1em plus 0.5em minus 0.4em\relax New York, NY, USA: ACM,
  2009, pp. 531--540.

\bibitem{madejski11}
M.~Madejski, M.~Johnson, and S.~Bellovin, ``The failure of online social
  network privacy settings,'' Columbia University, Tech. Rep. CUCS-010-11,
  2011.

\bibitem{ali2014self}
A.~Ali-Eldin and J.~van~den Berg, ``A self-disclosure framework for social
  mobile applications,'' in \emph{Proceedings of the 2014 6th International
  Conference on New Technologies, Mobility and Security (NTMS)}.\hskip 1em plus
  0.5em minus 0.4em\relax IEEE, 2014, pp. 1--5.

\bibitem{singh2013norms}
M.~P. Singh, ``Norms as a basis for governing sociotechnical systems,''
  \emph{ACM Transactions on Intelligent Systems and Technology (TIST)}, vol.~5,
  no.~1, p.~21, 2013.

\bibitem{rovatsos2014multiagent}
M.~Rovatsos, ``Multiagent systems for social computation,'' in
  \emph{Proceedings of the 2014 international conference on Autonomous agents
  and multi-agent systems (AAMAS)}.\hskip 1em plus 0.5em minus 0.4em\relax
  IFAAMAS, 2014, pp. 1165--1168.

\bibitem{krupa2012handling}
Y.~Krupa and L.~Vercouter, ``Handling privacy as contextual integrity in
  decentralized virtual communities: The privacias framework,'' \emph{Web
  Intelligence and Agent Systems}, vol.~10, no.~1, pp. 105--116, 2012.

\bibitem{ciortea2012designing}
A.~Ciortea, Y.~Krupa, and L.~Vercouter, ``Designing privacy-aware social
  networks: a multi-agent approach,'' in \emph{Proceedings of the 2nd
  International Conference on Web Intelligence, Mining and Semantics}.\hskip
  1em plus 0.5em minus 0.4em\relax ACM, 2012, pp. 1--8.

\bibitem{kokciyancommitment}
N.~K{\"o}kciyan and P.~Yolum, ``Commitment-based privacy management in online
  social networks,'' in \emph{First International Workshop on the Multiagent
  Foundations of Social Computing ({MFSC@AAMAS-2014})}, 2014, pp. 1--12.

\bibitem{such12sddm}
J.~M. Such, A.~Espinosa, A.~Garc\'{i}a-Fornes, and C.~Sierra,
  ``{S}elf-disclosure {D}ecision {M}aking based on {I}ntimacy and {P}rivacy,''
  \emph{Information Sciences}, vol. 211, pp. 93--111, 2012.

\bibitem{wishart2010collaborative}
R.~Wishart, D.~Corapi, S.~Marinovic, and M.~Sloman, ``Collaborative privacy
  policy authoring in a social networking context,'' in \emph{Proceedings of
  the 2010 IEEE International Symposium on Policies for Distributed Systems and
  Networks (POLICY)}.\hskip 1em plus 0.5em minus 0.4em\relax IEEE, 2010, pp.
  1--8.

\bibitem{squicciarini2009collective}
A.~C. Squicciarini, M.~Shehab, and F.~Paci, ``Collective privacy management in
  social networks,'' in \emph{Proceedings of the 18th international conference
  on World wide web ({WWW})}.\hskip 1em plus 0.5em minus 0.4em\relax ACM, 2009,
  pp. 521--530.

\bibitem{hu2012detecting}
\BIBentryALTinterwordspacing
H.~Hu, G.-J. Ahn, and J.~Jorgensen, ``Detecting and resolving privacy conflicts
  for collaborative data sharing in online social networks,'' in
  \emph{Proceedings of the 27th Annual Computer Security Applications
  Conference ({ACSAC})}.\hskip 1em plus 0.5em minus 0.4em\relax New York, NY,
  USA: ACM, 2011, pp. 103--112. [Online]. Available:
  \url{http://doi.acm.org/10.1145/2076732.2076747}
\BIBentrySTDinterwordspacing

\bibitem{carminati2011collaborative}
B.~Carminati and E.~Ferrari, ``Collaborative access control in on-line social
  networks,'' in \emph{Collaborative Computing: Networking, Applications and
  Worksharing (CollaborateCom), 2011 7th International Conference on}.\hskip
  1em plus 0.5em minus 0.4em\relax IEEE, 2011, pp. 231--240.

\bibitem{thomas2010unfriendly}
K.~Thomas, C.~Grier, and D.~M. Nicol, ``unfriendly: Multi-party privacy risks
  in social networks,'' in \emph{Privacy Enhancing Technologies}.\hskip 1em
  plus 0.5em minus 0.4em\relax Springer, 2010, pp. 236--252.

\bibitem{hu2012multiparty}
H.~Hu, G.~Ahn, and J.~Jorgensen, ``Multiparty access control for online social
  networks: model and mechanisms,'' \emph{IEEE Transactions on Knowledge and
  Data Engineering}, 2012.

\bibitem{sierra07}
C.~Sierra and J.~Debenham, ``The {LOGIC} negotiation model,'' in \emph{AAMAS
  '07: Proceedings of the 6th international joint conference on Autonomous
  agents and multiagent systems}.\hskip 1em plus 0.5em minus 0.4em\relax ACM,
  2007, pp. 1--8.

\end{thebibliography}

%\begin{IEEEbiographynophoto}{Jose M. Such}
%is Lecturer (Assistant Professor) in the School of Computing and Communications at Lancaster University (UK), which he joined in 2012. He was previously
%￼research fellow at Universitat Politecnica de Valencia (Spain), by which he was awarded a PhD in Computer Science in 2011. His main research interests are on the intersection between Artificial Intelligence and Cyber Security, and in particular, intelligent/automated approaches to Privacy, Identity Management, Access Control Models, Trust and Reputation. He is also interested in Human Factors in CyberSecurity and Machine learning applied to CyberSecurity.
%\begin{IEEEbiographynophoto}{Michael Rovatsos}
%is Senior Lecturer (Associate Professor) and Director of the Centre for Intelligent Systems and their Applications at the School of Informatics of the University of Edinburgh, which he joined in 2004. His main research interests are in multiagent systems, and he has published over 60 papers in this area with a particular focus on agent communication, multiagent learning, and multiagent planning. Michael holds a PhD from the Technical University of Munich (2004) and a Diploma from the University of Saarbruecken (1999), both in Computer Science.

\end{document}